\documentclass[11pt,DIV12]{scrartcl}

\usepackage{graphicx}
\usepackage{amsmath,amssymb}
\usepackage{euscript}
\usepackage{a4wide}
\usepackage{booktabs,dcolumn,multirow,colortbl}
\usepackage[pdftex,table]{xcolor}
\usepackage{units}
\usepackage{slashed}

\newcommand{\beq}{\begin{equation}}
\newcommand{\eeq}{\end{equation}}
\newcommand{\bea}{\begin{eqnarray}}
\newcommand{\eea}{\end{eqnarray}}

\newcommand{\subsect}[1]{\subsection[#1]{\boldmath #1}}

\newcommand{\tto}{\!\to\!}

\newcommand{\gsim}{\lower.7ex\hbox{$
\;\stackrel{\textstyle>}{\sim}\;$}}
\newcommand{\lsim}{\lower.7ex\hbox{$
\;\stackrel{\textstyle<}{\sim}\;$}}

\newcommand{\bibit}[1]{\bibitem{#1}}

\newcommand{\aver}[1]{\langle #1\rangle}

\newcommand{\La}{\overline{\Lambda}}

\newcommand{\Lam}{\Lambda_{\rm QCD}}
\newcommand{\mhad}{\mu_{\rm hadr}}

\newcommand{\GeV}{\,\mbox{GeV}}
\newcommand{\MeV}{\,\mbox{MeV}}
\newcommand{\matel}[3]{\langle #1|#2|#3\rangle}
\newcommand{\state}[1]{|#1\rangle}

\newcommand{\eod}{\end{document}}

\newcommand{\msp}[1]{\mbox{\hspace*{#1mm}~}}

\begin{document}
\begin{titlepage}
\begin{flushright}
SI-HEP-2010-11 \\[0.2cm]
\end{flushright}

\vspace{2.5cm}
\begin{center}
{\Large\bf
Higher Order Power Corrections   in
Inclusive \boldmath  $B$  Decays} 
\end{center}

\vspace{0.5cm}
\begin{center}
{\sc Th.~Mannel, S.~Turczyk and N.~Uraltsev}\raisebox{4pt}{$^{a*}$} \\[2mm]
{Theoretische Physik 1, Fachbereich Physik, \\ 
Universit\"at Siegen.  D-57068 Siegen, Germany}\\[10pt]
\scalebox{.935}{\sf  ~$^a$\,{\tt also}\, Department of Physics,
University of Notre Dame, 
Notre Dame, IN 46556 \,USA}  
\end{center}

\vspace{0.8cm}
\begin{abstract}
\vspace{0.2cm}\noindent
We discuss order $1/m_b^4$ and $1/m_b^5$ corrections in inclusive semileptonic
decay of a $B$ meson. We identify relevant hadronic matrix elements of
dimension seven and eight and estimate them using the ground-state saturation
approximation. Within this approach the effects on the integrated rate and on
kinematic moments are estimated. The overall relative shift in $V_{cb}$ turns
out about $+0.4\%$ as applied to the existing fits. Similar estimates are
presented for $B\tto X_s+\gamma$ decays.
\end{abstract}

\vfill

\noindent
\hrulefill\hspace*{290pt} \\[1pt]
$^*$\,\scalebox{.894}{On leave of absence from Petersburg Nuclear
Physics Institute, Gatchina, St.\,Petersburg 188300, Russia}

\thispagestyle{empty}

\end{titlepage}

\addtocounter{page}{-1}

\newpage
\tableofcontents
\section{Introduction}
The operator product expansion (OPE) for heavy hadron decays has become the
standard tool for the evaluation of differential decay rates. While the
quality of the OPE may vary in different corners of the whole phase space, it
has been established that for sufficiently inclusive observables the OPE
yields an expansion in $\Lam / m_Q$ ($Q$ is beauty or charm) which, in case of
beauty weak decays, converges reasonably well, at least judging from the low
orders calculated up to now.  However, it has been argued that the OPE results
in an asymptotic series with limitations paralleling those for the
perturbative series. In particular, this implies a deteriorating behavior at
sufficiently high orders; therefore, it is well motivated to investigate the
higher orders of the power expansion.

The OPE for inclusive decays yields an expansion in $\Lam /m_b$ with the
coefficients which are themselves series in $\alpha_s$. Consequently we end up
with a double expansion in the two parameters. Currently the leading power
term, the partonic rate is known to order $\alpha_s^2$, including the
differential distributions relevant for the calculation of moments
\cite{melnik}.  The coefficients of even the first nonperturbative corrections
are not completely known to order $\alpha_s$; the one for the chromomagnetic
correction is only known at tree level, while the coefficient of the kinetic
operator \cite{becherboos} can be related to the leading power term by
considering the OPE for a moving $B$ meson.\footnote{This relation is
  sometimes referred to as a reparametrization invariance.}  Higher order
terms in the OPE are only known at tree level where they can be directly
constructed by the method discussed in \cite{1m4}. Certain enhanced terms to
order $1/m_Q^4$ were calculated for the total decay rate to order ${\cal
  O}(\alpha_s)$ \cite{ic}.

The present paper focuses on the higher orders in the $1/m_b$ expansion for $b
\tto c$ inclusive semileptonic decay at tree level. As expected, we observe a
proliferation of nonperturbative expectation values starting from $1/m_b^4$.
Nevertheless, we identify the set of hadronic parameters through order
$1/m_b^5$, all those that appear at tree level.

Since at higher orders the number of nonperturbative parameters becomes too
large, a straightforward fit to the data to extract them is not possible.  To
get around this obstacle, we suggest a simple way to estimate the required
expectation values.  This ``ground state saturation assumption'' evaluates
higher-dimension matrix elements in terms of $\mu_\pi^2$, $\mu_G^2$,
$\rho_D^3$ and $\rho_{LS}^3$.  Using the reasonably well known values for the
latter from the data we are in the position to numerically assess the
higher-order parameters in the power expansion.

In the next section we describe a general method to calculate the higher
orders in the $1/m_Q$ expansion at tree level. In Sect.~\ref{m4} we apply this
to order $1/m_b^4$, including the set of the nonperturbative expectation
values, and in section~\ref{m5} we do the same for the terms at order
$1/m_b^5$. In Sect.~\ref{GSA} we set up the factorization approximation for
the heavy quark matrix elements and apply it to evaluate the $B$-meson
expectation values of dimension seven and eight operators appearing at
$1/m_b^4$ and $1/m_b^5$.  Using these results we perform a pilot numerical
study of the corrections at order $1/m_b^4$ and $1/m_b^5$ to the hadronic and
leptonic moments and to $|V_{cb}|$ in Sect.~\ref{num}.  The impact of
higher-order corrections turns out sizable in general; in particular, the
extracted value of the Darwin term may shift significantly.  Nevertheless, the
value of $|V_{cb}|$ remains stable.  Sect.~\ref{bsg} reports a brief summary
of the similar analysis for the $B\tto X_s+\gamma$ decays, including the
analytic expressions for the corrections to the rate and the photon energy
moments. Here the impact of $1/m_b^4$ and $1/m_b^5$ terms appears to be
insignificant.  Sect.~\ref{concl} summarizes our results and presents the main
conclusions.  Certain aspects of the derivation related to the factorization
approximation are relegated to Appendices.

\section{Calculational Scheme for \boldmath $1/m_b^n$ \unboldmath at Tree 
Level} \label{gen}

In this section we give a brief summary of our method to calculate power
corrections. It follows the calculational scheme used in
Ref.~\cite{1m4}.  The starting point for the
calculation is the differential decay rate 
\begin{equation}
\text{d}\Gamma = 
\frac{G_F^2 |V_{cb}|^2}{2} W_{\mu\nu}\, {\rm d}L^{\mu \nu},
\end{equation}
where the leptonic tensor ${\rm d} L^{\mu \nu}$ contains
all information on the leptons including their phase space element.
The nontrivial nonperturbative dynamics is encoded in the
hadronic tensor $W_{\mu\nu}$ related, by the optical theorem,
to the discontinuity of the time-ordered product of two weak currents across the
cut.  One then starts with a correlator of two hadronic currents 
\begin{equation} 
\label{Tprod}
  T_{\mu \nu}(q) = \frac{1}{2M_B} \int \text{d}^4 x\,  
e^{-i q x} \langle B(P) | iT\: \overline{b}(x) 
\Gamma^\dagger_\nu c(x) \:\overline{c}(0)\Gamma_\mu b(0) | B(P) \rangle\, .
\end{equation}
Here 
\begin{equation}  
\label{smcurrent}
\Gamma_\mu = \gamma_\mu(1-\gamma_5) 
\end{equation} 
is the left-handed weak current and $q$ the momentum transfer to the leptons while
$P$ denotes the $B$ meson momentum, and
\begin{equation} 
\label{OT}
 W_{\mu\nu} = 2\,\text{Im}\, T_{\mu \nu}\, . 
\end{equation}

Two-index amplitude $T_{\mu\nu}(q)$ or its discontinuity  $W_{\mu\nu}(q)$ can be
decomposed into five tensor structures with the coefficients given by scalar
functions of $q_0 = v \cdot q$ and $q^2$:
\bea  
W_{\mu\nu}(q) &=& -w_1(q_0, q^2) g_{\mu\nu} +w_2(q_0, q^2) v_\mu v_\nu 
-iw_3(q_0, q^2) \epsilon_{\mu\nu\rho\lambda} v^\rho q^\lambda  \label{w1w5} 
\\ \nonumber && +
w_4(q_0, q^2) q_\mu q_\nu + w_5(q_0, q^2) (q_\mu v_\nu+v_\mu q_\nu) ,
\eea
where $v = P/M_B$ is the four-velocity of the decaying $B$ meson. 
The structure functions $w_4$ and $w_5$ do not affect decay amplitudes with
massless leptons, and the fully differential semileptonic rate
depending on the charge lepton energy $E_\ell$ is
given by 
\bea
\nonumber
\frac{{\rm d}^3 \Gamma}{{\rm d}E_{\ell\,}  {\rm d}q^{2\,} {\rm d}q_0 }
&\msp{-4}= \msp{-4}& \frac{G_F^2 |V_{cb}|^2}{32\pi^4}\,
\vartheta\!\left(q_0\!-\!E_\ell\!-\!\mbox{$\frac{q^2}{4E_\ell}$}\right)
\vartheta(E_\ell) 
\,\vartheta(q^2) \;\times \\ 
& & \msp{20} \left\{
2 q^2 w_1+[4E_\ell (q_0\!-\!E_\ell)\!-\!q^2]w_2 +
2q^2(2E_\ell\!-\!q_0) w_3
\right\} .
\msp{20}\label{doctor92}\msp{-20}
\eea

\begin{figure}[hptb]
	\centering\includegraphics{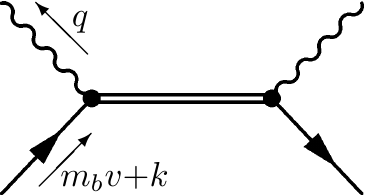}
	\caption{Tree level Feynman diagram for the hadronic tensor in 
inclusive semileptonic $B$ decays}
	\label{fig1}
\end{figure}

The tree-level expansion in $1/m_b$ is most easily set up by looking at the Feynman  
diagram in Fig.~\ref{fig1}. The double line denotes the Green function of the 
charm quark propagating in the background field of soft gluons in  
the $B$ meson. We `rephase' the $b$ fields according to $b(x)\tto e^{-im_b (vx)}
b(x)$; this makes the actual $b$-quark momentum operator to become
\begin{equation} 
\label{bquarkmom}
	p_b = m_b  v + i D .
\end{equation}
The phase factor from $\bar{b}(x)$ in Eq.~(\ref{Tprod}) combines with the
background-field $c$-quark propagator \cite{banda} 
$1/(i\slashed{D}\!-\!m_c)$ to yield charm propagator
\begin{equation} \label{BGFprop}
	S_{\rm BGF} = \frac{1}{\slashed{p}+ i \slashed{D} -m_c} , \qquad 
p_\mu\equiv m_bv_\mu\!-\!q_\mu
\end{equation}
to be inserted between the nonrelativistic `rephased' $b$ fields;
this corresponds to the $b$ quark momentum Eq.~(\ref{bquarkmom}).

For semileptonic processes at tree level we only need to multiply
(\ref{BGFprop}) by the appropriate Dirac matrices (\ref{smcurrent})
for the left handed current.  A calculation of the OPE series to order $1/m_b^n$
requires to expand this expression to $n^{th}$ order in the covariant
derivative $(iD)$ according to
\begin{equation} 
\label{BGFexpansion}
S_{\rm BGF} = \frac{1}{\slashed{p}\!-\!m_c} +  \frac{1}{\slashed{p}\!-\!m_c} 
( -i \slashed{D}) \frac{1}{\slashed{p}\!-\!m_c}+ \frac{1}{\slashed{p}\!-\!m_c} 
( -i \slashed{D}) \frac{1}{\slashed{p}\!-\!m_c} ( -i \slashed{D}) 
\frac{1}{\slashed{p}\!-\!m_c} + \cdots 
\end{equation}
The covariant derivatives playing the role of the `residual' momentum, do not
commute in general; the above expansion takes care of their
ordering. Calculating structure functions we take discontinuity of 
$S_{\rm BGF}$; successive terms in the expansion yield higher derivatives of
$\delta(m_b^2\!-\!m_c^2\!+\!q^2\!-\!2m_b q_0)$, together with 
powers  of the spinor factor 
$m_b\slashed v\!-\!\slashed{q}\!+\!m_c$ ($m_c$ term can be consistently
dropped for purely left-handed weak vertices) producing polynomials 
in $q$. The cumbersome, for high-order
terms, general product of many Dirac matrices is simplified using a computer 
routine. 

Finally, the thus  obtained operator expansion should be supplemented by 
the $B$-meson expectation values of the general form 
\begin{equation}
	\bar{b}_{\alpha} (iD_{\mu_1}) ....  (iD_{\mu_n}) b_{\beta}  
\end{equation} 
where the spinor indices of the $b$ fields are shown explicitly.  The
field is the full four-component QCD field yet redefined by a phase factor.
Note that performing the OPE this way yields only local operators; however,
their expectation values still contain a nontrivial mass dependence from the
small `lower' bispinor components and from the explicit subleading terms in
the equations of motions for $b(x)$ that will be discussed in the following,
as well as from the fact that the $B$ meson states are the eigenstate of the
finite-mass Hamiltonian.

Selecting the basis for independent expectation values to different orders in
$1/m_b$ may be accomplished in different ways by partially reshuffling the
power-suppressed pieces into the definition of the lower-dimension expectation
values. The most familiar example through order $\mhad^3$ is the
chromomagnetic expectation value $\mu_G^2$ which is usually defined as
$\aver{\bar{b} (-\vec\sigma \vec B) b}$, yet often the full Lorentz-scalar
operator $\bar{b} \frac{i}{2}\sigma_{\mu\nu} G^{\mu\nu} b$ is used instead;
these differ explicitly by $(\rho_D^3\!+\!\rho^3_{LS})/2m_b$. One common
strategy assumes that the lower components of the spinors are explicitly
excluded (while passing to the genuine nonrelativistic $b$ fields, i.e.\ to
the upper components corrected by power terms, still may or may not be
performed).  In this study we follow an alternative scheme used in
Ref.~\cite{1m4} which may not necessary parallel such a separation; it
generally keeps all four components of the $b$ field and only reduces the
number of operators applying equations of motion. While in this scheme the
generic operators are not expressed through the conventional nonrelativistic
$b$ fields, it has an advantage of being more compact and more native to the
tree-level local OPE calculations.

The enumeration and evaluation of these matrix elements is conveniently performed  
in a recursive fashion, starting with the operators of the 
highest dimension, i.e.\ those with the maximal number of 
covariant derivatives. There one can neglect all
$1/m_b$-suppressed pieces and consider them in the static limit; to facilitate
passing on to the next step we, however, still use the same full QCD fields. 
In the static limit one has two possible spinor structures of opposite parity 
\cite{zerec}: 
\beq
\matel{B} {\bar b_{\alpha} (iD_{\mu_1}) ....  
(iD_{\mu_n}) b_{\beta}} {B} = \left(\frac{1\!+\!\slashed{v}}{2}\right)_{\beta
\alpha} A_{\mu_1 \mu_2 \cdots
\mu_n} + \left(s_\lambda\right)_{\beta
\alpha} B_{\mu_1 \mu_2 \cdots \mu_n}^\lambda 
\label{s9}
\eeq
where $\frac{1\!+\!\slashed{v}}{2}$ and  $s_\lambda =  
\frac{1\!+\!\slashed{v}}{2}\gamma_\lambda \gamma_5
\frac{1\!+\!\slashed{v}}{2}$ are the generalizations, 
to the case of arbitrary velocity
frame, of the unit and Pauli matrices projected onto the `upper'
components. Temporarily introducing an arbitrary restframe velocity vector 
$v_\mu$, the standard tensor decomposition technique expresses 
the tensor structures $A$ and $B$ through a minimal set of fundamental expectation
values at a given order. This can be done by contracting the indices in 
all possible ways. In the following, the matrix elements of this set are 
referred to as 
{\sl basic parameters} for a certain order in $1/m_b$. We emphasize that such
relations are purely algebraic as long as a general parameterization like in
Eq.~(\ref{s9}) is adopted; they do not imply any assumption about heavy
quark limit or subleading correction, or about dynamics -- they are just the
result of rotational symmetry of the $B$ expectation values. Consequently, the
same way to relate the corresponding tensors to the basic parameters can be
used for the expectation values of the operators of any dimension. However,
the basic parameters encountered at lower dimension generally include
higher-order pieces which depend on the choice of the basic set.

For the operators of dimensions lower than the highest, therefore, we have to 
include in consideration originally all possible 16 bilinear bispinor
structures including power-suppressed ones when 
generalizing Eq.~(\ref{s9}). Likewise,
we use equations of motions for the $b$ field to relate certain structures 
to higher-dimension expectation values; we do this at the second step, at the
level of invariants thus reducing directly the set of basic parameters. 
In particular, we successively replace the rightmost and
leftmost time derivatives $(ivD)$ acting on $b$ ($\bar b$) according to the 
QCD equation
of motion (see Ref.~\cite{optical}, Sect.~2); the same is done to reduce 
certain bispinor structures: 
\beq
(ivD) b = - \frac{(iD)^2 +\frac{1}{2}\sigma_{\mu\nu}G^{\mu\nu}}{2m_b} \,b, \qquad 
\frac{1\!-\! \slashed{v}}{2} b=  \frac{i\slashed{D}}{2m_b}b\, , \qquad 
 \gamma_\mu \gamma_\nu = g_{\mu\nu} - i \sigma_{\mu\nu}  
\label{eqmot}
\eeq
where we have also specified the convention used for 
$\sigma_{\mu\nu}$.\footnote{Note that this convention differs from the one
  adopted in a number of earlier papers on the subject.}
Using various identities for the Dirac matrices combined with the second of
the above equations one can reduce the number of spin structures to one
spin-singlet and one spin-triplet, cf.\ Eq.~(\ref{s9}); for the latter 
we choose the dual spin matrix $-i\sigma_{\mu\nu}$. The spin reduction is less
automated than eliminating time derivatives of the $b$ fields, nevertheless to
the order we are interested in, through operators with five derivatives, this is
straightforward. The reduction algorithm includes effectively eliminating 
the terms which would mix lower and upper components of the spinors; then the
leading order of a given expectation value is directly given by the number of
covariant derivatives in the operator. Spin-triplet operators contain
four-dimensional $-i\sigma_{\mu\nu}$ (reduced to three-dimensional 
$\sigma_{mn}$ to the leading order), and spin-singlet ones are free from any Dirac
structures. The added convenience of such bases is that the corrections 
from the presence of 
lower components are automatically $1/m_b^2$ suppressed. However, we bear in
mind that the classification itself upon spin properties is valid only to the
leading order for a given operator. Upon completing the reduction we end up
with the expression in terms of basic set at each order in $\Lam$; the 
expectation values at a given order generally add $1/m_b$-suppressed terms
proportional to the basic expectation values of higher orders.

For the meson expectation values Lorentz covariance of the operators proper is
undermined by presence of the external $b$-states selecting the preferred
frame; only rotation invariance remains. Therefore, the time and the spatial
derivatives (and indices in general) are physically distinct and should be
treated differently. Nevertheless, we write the operators in a superficially
Lorentz-invariant form employing the $b$-hadron velocity, as a remnant of the
way the relations to the basic set have been derived. We 
denote the projector onto the spatial components  by
\[ \Pi_{\mu \nu} = g_{\mu \nu} - v_\mu v_\nu . \]
Then the `usual' basic parameters, those appearing through order
$\Lam^3$ can be written as
\begin{align*}
\nonumber
2M_B\, \mu_\pi^2 &= - \langle B | \bar b_v \, i D_\rho i D_\sigma  \,b_v | 
B \rangle \,\,  \Pi^{\rho \sigma}  \\
\nonumber
2M_B\, \mu_G^2 &= \frac12 \langle B | \bar b_v \,\big[ i D_\rho ,  
iD_\lambda\big]  \big (-i \sigma_{\alpha \beta}\big)\,b_v | B \rangle \,\,  
\Pi^{\alpha \rho} \Pi^{\beta \lambda} \\
\end{align*}
\begin{align}
\nonumber
2M_B\, \rho_D^3 &= \frac12 \langle B | \bar b_v \,\Big[ i D_\rho  , 
\big[i D_\sigma ,  i D_\lambda \big]\Big]  \,b_v | B \rangle \,\,  
\Pi^{\rho \lambda} v^\sigma \\
2M_B\, \rho_{LS}^3 &= \frac12 \langle B | \bar b_v \, 
\Big \lbrace i D_\rho, \big[i D_\sigma ,  i D_\lambda \big]\Big\rbrace  
\big (-i \sigma_{\alpha \beta}\big)\,b_v | B \rangle \,\,  \Pi^{\alpha \rho} 
\Pi^{\beta \lambda} v^\sigma 
\label{m23}
\end{align}
where full QCD $b$ fields and states are assumed. Note that although these
parameters may differ in higher-order $1/m_b$ pieces from the traditionally
used four expectation values, the difference for $\mu_\pi^2$ and $\mu_G^2$
is only $1/m_b^2$.

Performing the calculation along the lines described above and contracting the
indices obtained in the OPE series Eq.~(\ref{BGFexpansion}) with the tensor
structures derived for the expectation values generalizing Eq.~(\ref{s9}) we
obtain the formal $1/m_b$ expansion of the $B$-decay structure functions, in
this case for semileptonic decays. Using the basic parameters in
Eq.~(\ref{m23}) we readily reproduce the known results for the structure
functions, lepton spectrum and the total decay rate
through order $1/m_b^3$ \cite{koyrakh,grekap}. 

\subsect{Expectation values at $1/m_b^4$ } \label{m4} 

Passing to higher-order power correction requires first to find independent
expectation values which can form the set of basic parameters to this order.
The starting point are all possible index contractions of a matrix element of
the form
\begin{equation}    
 \langle B | \bar b_v \, i D_\rho i D_\sigma i D_\lambda i D_\delta 
\Gamma \,b_v | B \rangle \, , 
\end{equation} 
where $\Gamma$ is either the unit matrix or $-i\sigma_{\mu \nu}$. 
However, not all of them can appear  
independently;  $T$-invariance of the correlator (\ref{Tprod}) ensures that
only Hermitian operators are present. This also automatically yields a real
differential rate, since the coefficient functions are real. 

Taking this into account, 
we end up with nine dimension-$7$ operators, four spin-singlet and five spin-triplet:
\begin{align}
\nonumber
2M_B\, m_1 &= \langle B | \bar b_v \, i D_\rho i D_\sigma i D_\lambda iD_\delta 
\,b_v | B \rangle\,\, \scalebox{1.1}{$\frac{1}{3}$} 
\left(\Pi^{\rho \sigma}\Pi^{\lambda \delta} + 
\Pi^{\rho \lambda}\Pi^{\sigma \delta}+\Pi^{\rho \delta}\Pi^{\sigma \lambda}\right) \\
\nonumber
2M_B\, m_2 &= \langle B | \bar b_v \, \big[ i D_\rho ,  i D_\sigma \big] 
\big[ i D_\lambda ,  i D_\delta\big] \,b_v | B \rangle \:\Pi^{\rho\delta}v^\sigma 
v^\lambda \\
\nonumber
2M_B\, m_3 &= \langle B | \bar b_v \, \big[i D_\rho ,  i D_\sigma\big]
\big[ iD_\lambda, iD_\delta\big] \,b_v|B\rangle\:\Pi^{\rho \lambda}\Pi^{\sigma \delta} \\
\nonumber
2M_B\, m_4 &= \langle B | \bar b_v \, \Big\lbrace i D_\rho , \Big[i D_\sigma ,
\big[ i D_\lambda , i D_\delta\big]\Big]\Big \rbrace \,b_v | B \rangle \:
\Pi^{\sigma \lambda}\Pi^{\rho \delta} \rule[-10pt]{0pt}{8pt}\\
\nonumber
2M_B\, m_{5} &= \langle B | \bar b_v \, \big[ i D_\rho  , i D_\sigma\big] 
\big[ i D_\lambda ,  i D_\delta\big]  \big (-i \sigma_{\alpha \beta}\big)\,b_v
| B \rangle \,\,  \Pi^{\alpha \rho} \Pi^{\beta \delta} v^\sigma v^\lambda \\
\nonumber
2M_B\, m_6 &= \langle B | \bar b_v \, \big[ i D_\rho  , i D_\sigma \big] 
\big[i D_\lambda , i D_\delta\big]  \big (-i \sigma_{\alpha \beta}\big)\,b_v 
| B \rangle \,\, \Pi^{\alpha \sigma} \Pi^{\beta \lambda} \Pi^{\rho \delta} \\
\nonumber
2M_B\, m_7 &= \langle B | \bar b_v \,\Big \lbrace \big \lbrace i D_\rho,  
i D_\sigma \big \rbrace ,\big[ i D_\lambda , i D_\delta \big] \Big\rbrace 
\big (-i \sigma_{\alpha \beta}\big)\,b_v | B \rangle \,\, \Pi^{\sigma \lambda}
\Pi^{\alpha \rho} \Pi^{\beta \delta} \\
\nonumber
2M_B\, m_8 &= \langle B | \bar b_v \,\Big \lbrace \big \lbrace
i D_\rho ,  i D_\sigma \big \rbrace, \big[i D_\lambda  , i D_\delta \big] 
\Big\rbrace \big (-i \sigma_{\alpha \beta}\big)\,b_v | B \rangle \: 
\Pi^{\rho \sigma } \Pi^{\alpha \lambda} \Pi^{\beta \delta} \\
2M_B\, m_{9} &= \langle B | \bar b_v \, \bigg[ i D_\rho ,  \Big[ i D_\sigma ,
\big[ i D_\lambda ,  i D_\delta \big]\Big]\bigg]  \big (-i \sigma_{\alpha \beta}\big) 
\,b_v | B \rangle \,\, \Pi^{\rho \beta} \Pi^{\lambda \alpha} \Pi^{\sigma \delta}\,.
\end{align}

The commutator of covariant derivatives equals field strength, and all
operators above but the first one include as a factor an explicit gluon field
operator. Therefore, their expectation values without at least one gluon would
vanish, whether the hadronic state has zero momentum or not. Boosting the
$b$-hadron would not generate per se such expectation values. Only $m_1$ can
emerge from the boost; it can be identified with average $(\vec
p^{\,2})^2\!\equiv\!  \vec p ^{\,4}$ in the nonrelativistic picture. However,
a subtlety must be born in mind when pursuing such an interpretation, for
$\mu_\pi^2$ as defined in Eqs.~(\ref{m23}) contains, along with $\aver{\vec
  p^{\,2}}$, an extra piece $\propto \aver{\vec p ^{\,4}}/m_b^2$ which has to
be included when checking the effect of the boost through the OPE. This
$1/m_b^2$ piece comes both from the presence of the lower components of the
$b$ fields in $\mu_\pi^2$ in Eqs.~(\ref{m23}) and from the fact that the upper
components proper should be corrected at order $1/m_b^2\,$ by the additional
operator $1\!+\!\frac{(\vec \sigma i\vec D)^2}{8m_b^2} \!=\!
1\!+\!\frac{(i\vec D)^2+\vec\sigma \vec B}{8m_b^2}$ to represent the true
nonrelativistic fields \cite{optical}.

The corresponding correction most simply can be recovered using two identities
\beq
\frac{1}{2M_B}\,\matel{B}{\bar b \gamma_0 b}{B}=1, \qquad 
\frac{1}{2M_B}\,\matel{B}{\bar b  b}{B}=
1-\frac{1}{2m_b^2} (\mu_\pi^2-\mu_G^2) + \ldots\,;
\label{ident}
\eeq
this yields 
\beq
\mu_\pi^2 \Longleftrightarrow  \aver{\vec p ^{\,2}}-\frac{1}{2m_b^2}\aver{\vec
  p ^{\,4}} +{\cal O}\left(\frac{G_{\alpha\beta}}{m_b^2}, \frac{p^6}{m_b^4}\right).
\label{p4}
\eeq

The other operators can also be interpreted in a simple manner. The
expectation value $m_2$ is proportional to $\langle \bar{b}\, \vec{E}^2\, b
\rangle$; likewise is $m_3$ proportional to $\langle \bar{b}\, \vec{B}^2 \,b
\rangle$, where $\vec{E}$ and $\vec{B}$ are the chromoelectric and
chromomagnetic fields. The parameter $m_4$ is related to $\langle \vec{p}\!
\cdot\! {\rm rot }\, \vec{B} \rangle$ or, by the gluon field equations of
motion, to a combination of $m_2$ and $\aver{g_s^2\,
  \bar{b}\,(\vec{p}\vec{J})\, b}$ where ${\rm rot}$ is the non-Abelian version
of the rotator and $J_\mu$ the color octet flavor-singlet vector current of
light quarks.

The spin-dependent operators have similar interpretation: $m_5 \propto \langle
\bar{b}\,\vec{s} \!\cdot\! \vec{E} \!\times\! \vec{E}\, b\rangle$ and $m_6
\propto \langle \bar{b}\,\vec{s} \!\cdot\! \vec{B}\! \times\! \vec{B}\, b
\rangle$ (different components of the chromoelectric and chromomagnetic fields
do not commute) while $m_7 \propto \aver{(\vec{p})^2(\vec{s} \!\cdot
  \!\vec{B})\!-\!  (\vec{s} \!\cdot \!\vec{p}) (\vec{p} \!\cdot\! \vec{B})}$,
$m_8 \propto \langle (\vec{p})^2 (\vec{s} \!\cdot \!\vec{B}) \rangle$ and
$m_9$ is a combination of $\aver{\Delta (\vec{s} \!\cdot\! \vec{B})}$ and
$\aver{\vec{s} \!\cdot\! \vec{B}\! \times\! \vec{B}}$.

\subsect{Expectation values at $1/m_b^5$ } \label{m5}

The number of possible operators with five covariant derivatives and of their
expectation values further increases. Requiring $T$-invariance we get 
18 new parameters in total, 7 of which are spin singlet 
and 11 are spin triplet:
\begin{align*}
\nonumber
2M_B r_1 &= \langle B | \bar b \,i  D_\rho\, (i v \cdot D)^3\, i  D^\rho \, b | B \rangle \\
\nonumber 
2M_B r_2 &= \langle B | \bar b \,i  D_\rho\, (i v \cdot D)\, i  D^\rho\, i  D_\sigma\, i  D^\sigma \, b | B \rangle \\
\nonumber
2M_B r_3 &= \langle B | \bar b \,i  D_\rho\, (i v \cdot D)\, i  D_\sigma\, i
D^\rho\, i  D^\sigma \, b | B \rangle \\  
\nonumber
2M_B r_4 &= \langle B | \bar b \,i  D_\rho\, (i v \cdot D)\, i  D_\sigma\, i
D^\sigma\, i  D^\rho \, b | B \rangle \\  
\nonumber
2M_B r_5 &= \langle B | \bar b \,i  D_\rho\, i  D^\rho\,(i v \cdot D)\,  i
D_\sigma\, i  D^\sigma \, b | B \rangle \\  
\nonumber
2M_B r_6 &= \langle B | \bar b \,i  D_\rho\, i  D_\sigma\, (i v \cdot D)\, i
D^\sigma\, i  D^\rho \, b | B \rangle \\  
\nonumber
2M_B r_7 &= \langle B | \bar b \,i  D_\rho\, i  D_\sigma\, (i v \cdot D)\, i
D^\rho\, i  D^\sigma \, b | B \rangle \rule[-10pt]{0pt}{8pt}\\  
\nonumber
2M_B r_{8} &= \langle B | \bar b \,i   D_\mu \, (i v \cdot D)^3\, i   D_\nu
\, (-i \sigma^{\mu \nu })\,b | B \rangle \\  
\nonumber
2M_B r_{9} &= \langle B | \bar b \,i   D_\mu \, (i v \cdot D)\, i   D_\nu
\, i   D_\rho\, i   D^\rho \,(-i \sigma^{\mu \nu })\, b | B \rangle \\  
\nonumber
2M_B r_{10} &= \langle B | \bar b \,i   D_\rho\, (i v \cdot D)\, i
D^\rho\, i   D_\mu \, i   D_\nu  \,(-i \sigma^{\mu \nu })\, b | B \rangle \\
\nonumber
2M_B r_{11} &= \langle B | \bar b \,i   D_\rho\, (i v \cdot D)\, i   D_\mu
\, i   D^\rho\, i   D_\nu  \,(-i \sigma^{\mu \nu })\, b | B \rangle \\  
\nonumber
2M_B r_{12} &= \langle B | \bar b \,i   D_\mu \, (i v \cdot D)\, i
D_\rho\, i   D_\nu \, i   D^\rho \,(-i \sigma^{\mu \nu })\, b | B \rangle \\
\end{align*}
\begin{align}
\nonumber
2M_B r_{13} &= \langle B | \bar b \,i   D_\rho\, (i v \cdot D)\, i   D_\mu
\, i   D_\nu \, i   D^\rho \,(-i \sigma^{\mu \nu })\, b | B \rangle \\  
\nonumber
2M_B r_{14} &= \langle B | \bar b \,i   D_\mu \, (i v \cdot D)\, i
D_\rho\, i   D^\rho\, i   D_\nu  \,(-i \sigma^{\mu \nu })\, b | B \rangle \\
\nonumber
2M_B r_{15} &= \langle B | \bar b \,i   D_\mu \, i   D_\nu \, (i v \cdot
D)\, i   D_\rho\, i   D^\rho \,(-i \sigma^{\mu \nu })\, b | B \rangle \\  
\nonumber
2M_B r_{16} &= \langle B | \bar b \,i   D_\rho\, i   D_\mu \, (i v \cdot
D)\, i   D_\nu \, i   D^\rho \,(-i \sigma^{\mu \nu })\, b | B \rangle \\  
\nonumber
2M_B r_{17} &= \langle B | \bar b \,i   D_\mu \, i   D_\rho\, (i v \cdot
D)\, i   D^\rho\, i   D_\nu  \,(-i \sigma^{\mu \nu })\, b | B \rangle \\  
2M_B r_{18} &= \langle B | \bar b \,i   D_\rho\, i   D_\mu \, (i v \cdot D)\, i   D^\rho\, i   D_\nu  \,(-i \sigma^{\mu \nu })\, b | B \rangle \,.
\end{align}
Note that at this dimension we have an odd number of derivatives, therefore at
least one time derivative $(ivD)$ must be present. This derivative can be
commuted through the rest until it acts on one of the $b$ quark operators,
which yields zero to this order. The remaining commutators can be expressed in
terms of chromoelectric field, so all these matrix elements are necessarily
proportional to the gluon field. Unlike the case of an even number of
derivatives, none of such expectation values can be generated by a $B$ boost
alone.

\subsect{OPE to orders $1/m_b^4$ and $1/m_b^5$}

Using the above set of expectation values and the expansion procedure
described earlier in this section, we derive all five semileptonic
decay structure functions of $B$ mesons through order $1/m_b^5$. They contain
$\delta(m_b^2\!-\!m_c^2\!+\!q^2\!-\!2m_bq_0)$ and its derivatives from first
through fifth. In view of extremely large number of lengthy terms due to
polynomial coefficients in $q_0$ and $q^2$, to different multiple derivatives
of the $\delta$-function for each of the expectation values and to algebraic
expressions resulting from reduction of various Lorentz structures in the
generic matrix elements to the basic parameters, we do not attempt to present
them explicitly. They are handled in Mathematica, with the typical size of the
file with their definition around $0.4\,{\rm MB}$.

The actual inclusive observables we are interested in involve kinematic
integrations which eliminate the $\delta$-function and its derivatives and
yield literal $1/m$ expansion for the effects associated with
higher-dimension expectation values. We obtain the particular observable
using Eq.~(\ref{doctor92}) and include the cut on lepton energy if
imposed. This is likewise done in Mathematica; for the moments with a cut we
evaluate the expressions numerically assuming certain values for the $D\!=\!8$
expectation values. 
Where hadronic mass moments are considered, we use the expression for $M_X^2$
\bea
M_X^2 \msp{-4}&=& \msp{-4} (m_b^2 - m_c^2 + q^2 - 2m_bq_0) + 2 (M_B - m_b)(m_b - q_0) +
(M_B - m_b)^2 
\nonumber  \\ 
 \msp{-4}& \equiv & \msp{-4} m_x^2+2(M_B - m_b) e_x+(M_B - m_b)^2;
\label{mx2}
\eea
the corresponding factor then multiplies the right hand side of
Eq.~(\ref{doctor92}). For integer $M_X^2$ moments the factor is a polynomial
in $M_B\!-\!m_b$ with the coefficients that are polynomials in $q_0$ and
$q^2$. The quantities $m_x^2$ and $e_x$ have a meaning of hadron invariant
mass squared and hadron energy, respectively, at the partonic level in the
decay of an isolated $b$ quark.

\section{Estimate of the hadronic expectation values} \label{GSA} 
As we have seen in the previous section, the independent hadronic matrix
elements start to proliferate at order $1/m_b^4$.  This would make their
complete extraction from data difficult if not impossible.  However, not all
the corrections may be expected equally important, and neglecting numerically
insignificant effects reduces the number of new parameters, possibly even to a
manageable level. Yet this requires estimating the scale of the emerging
expectation values well beyond naive dimensional guessing.

In the present paper we are concerned with the heavy quark expectation values
at tree level, appearing through dimension $8$, and in the following we
present a method for estimating them based on the saturation by intermediate
states.

Our approach starts with representing the matrix elements of the composite
operators of interest as a sum over the full set of zero-recoil intermediate
single-heavy--quark hadronic states.  While this is still exact, the
approximation scheme we shall employ is to retain in the sum the contribution
of only the multiplet of the allowed lowest-energy intermediate states. The
quality of this approximation depends on the degree of saturation of the sum
by the lowest states. The related accuracy in certain cases can be estimated
and the whole approximation refined.

\subsect{Intermediate state representation for the relevant 
matrix elements} \label{interm}

Our goal are forward matrix  elements of the form 
\begin{equation} 
\matel{B} {\bar{b} \; iD_{\mu_1} iD_{\mu_2} \cdots iD_{\mu_n} \,\Gamma \, b(0)
}{B}, 
\label{ME}
\end{equation}
where $\Gamma$ denotes an arbitrary Dirac matrix.
The representation is obtained by splitting the full 
chain $iD_{\mu_1} iD_{\mu_2} \cdots iD_{\mu_n}$ into 
$A\!=\!iD_{\mu_1} iD_{\mu_2} \cdots iD_{\mu_k}$ and 
$C\!=\!iD_{\mu_{k+1}} iD_{\mu_{k+2}} \cdots iD_{\mu_n}$.

The intermediate states representation says that 
\beq
\matel{B} {\bar{b} \; A \, C \,\Gamma \, b(0)
}{B} = \frac{1}{2M_B} \sum_n \; \matel{B} {\bar{b} \, A \,b(0)}{n} \cdot 
\matel{n} {\bar{b} \; C \,\Gamma \, b (0)}{B},
\label{s20}
\eeq
where we have assumed the $B$ mesons to be static and at rest, 
$\state{B}=\state{B(p\!=\!(M_B,\vec{0}))}$, and 
$\state{n}$ are the single-$b$ hadronic states with vanishing spatial
momentum.

We present an OPE-based proof of Eq.~(\ref{s20}); a more conventional
effective--field-theory derivation is given in Appendix A.  In either way we
introduce a ficticious heavy quark $Q$ which will be treated as static (i.e.,
$m_Q\tto \infty$ and normalization point is much lower than $m_Q$), and in the
OPE approach consider a correlator of the form
\beq 
T_{AC} (q_0) = \int {\rm d}^4 x \; e^{iq_0  x_0}
\matel{B}{i T \left\{  \bar{b}A Q(x)\, \bar{Q} C \Gamma b (0)\right\} }{B} ;
\label{s27} 
\eeq
note that spatial momentum transfer $\vec q\,$ has been explicitly set to vanish. 

We shall use the static limit for both $b$ and $Q$, and hence introduce the
`rephased' fields $\tilde Q(x)\!=\! e^{im_Q q_0 x_0}Q(x)$ and likewise for
$b$, and omit tilde in them in what follows.  The form of the resulting
exponent suggests to define $\omega \!=\! q_0 \!-\! m_b \!+\! m_Q$ as the
natural variable for $T_{AC}$, and $\frac{1}{2M_B}T_{AC}(\omega)$ is assumed
to have a heavy quark limit.

With large $m_Q$ we can perform the OPE for $T_{AC}(\omega)$ at $|\omega|
\!\gg\! \Lam$ still assuming that $|\omega| \!\ll\! m_Q$ and neglecting
thereby all powers of $1/m_Q$. In this case the propagator of $Q$ becomes
static,
\beq
iT \{ Q(x) \bar{Q}(0) \}  =\frac{1+\gamma_0}{2} \delta^3 (\vec{x}) \, \theta (x_0)\:
P\,\exp{\left( i\!\int_0^{x_0} \!\!\!A_0 \,{\rm d}x_0\right)},
\label{s29a}
\eeq
and yields 
\beq
T_{AC}(\omega)=
\matel{B}{\bar{b} \, A\frac{1}{-\omega\!-\!\pi_0\!-\!i0} C\:
\mbox{$\frac{1+\gamma_0}{2}$} \Gamma \, b }{B} \,,
\label{s30}
\eeq
where $\pi_0\!=\!iD_0$ is the time component of the covariant derivative.
This representation allows immediate expansion of $T_{AC}(\omega)$ in a series
in $1/\omega$ at large $|\omega|$:
\beq
T_{AC}(\omega) = -\sum_{k=0}^\infty\, \matel{B}{\bar{b} \, A
\frac{(-\pi_0)^k}{\omega^{k\!+\!1}} C\:
\mbox{$\frac{1+\gamma_0}{2}$} \Gamma \, b }{B}\,.
\label{s30a}
\eeq

Alternatively the scattering amplitude can be written through its
dispersion relation
\beq
T_{AC}(\omega) = \frac{1}{2 \pi i} \int_0^\infty  {\rm d}\epsilon \, 
\frac{1}{\epsilon \!-\! \omega \!+\! i 0} 
 \: \mbox{disc}\,T_{AC}(\epsilon)\, ,
\label{s31}
\eeq
where we have used the fact that in the static theory the scattering amplitude
has only one, `physical' cut corresponding to positive $\omega$.
The discontinuity is given by 
\beq
i \! \int {\rm d}^4 x \; e^{i\epsilon  x_0}
\matel{B}{\bar{b}A Q(x)\, \bar{Q} C \Gamma b (0) }{B} 
\label{s32}
\eeq
and amounts to
\beq
\mbox{disc}\,T_{AC}(\epsilon) =
\sum_{n_Q}\,
i \!\int {\rm d}^4 x \; e^{-i\vec{p}_n \vec x}\: e^{i(\epsilon\!-\!E_n) x_0}
\matel{B}{\bar{b}A Q(0)}{n_Q}\,\matel{n_Q}{ \bar{Q} C \Gamma b (0) }{B} ,
\label{s34}
\eeq
where the sum runs over the complete set of the intermediate states
$\state{n_Q}$; their overall spatial momentum is denoted by $\vec p_n$ and
energy by $E_n$. 

The spatial integration over ${\rm d}^3 x$ and integration over time  ${\rm d}
x_0$ in Eq.~(\ref{s34}) yield $(2\pi)^3 \delta^3(\vec p_n)$ and $2\pi
\,\delta(E_n\!-\!\epsilon)$, respectively. Therefore only the states with
vanishing spatial momentum are projected out, and we denote them as
$\state{n}$: 
\beq
\mbox{disc}\,T_{AC}(\epsilon) =
\sum_{n}
2\pi i\, \delta (\epsilon\!-\!E_n)\;
\matel{B}{\bar{b}A Q(0)}{n}\,\matel{n}{ \bar{Q} C \Gamma b (0) }{B} .
\label{s34a}
\eeq

Inserting the optical theorem relation (\ref{s34a}) into the dispersion integral 
(\ref{s31}) we get
\beq
T_{AC}(\omega) =
\sum_{n}
\frac{\matel{B}{\bar{b}A Q(0)}{n}\,\matel{n}{ \bar{Q} C \Gamma b (0) }{B}}
{E_n\!-\!\omega\!+\!i0 }\,,
\label{s34aa}
\eeq
and the large-$\omega$ expansion takes the form 
\beq
T_{AC}(\omega) = -\sum_{k\!=\!0}^\infty \, \frac{1}{\omega^{k\!+\!1}}\,
\sum_{n}\;
E_n^k\,\matel{B}{\bar{b}A Q(0)}{n}\,\matel{n}{ \bar{Q} C \Gamma b (0) }{B}\,.
\label{s34b}
\eeq

Equating the leading terms in $1/\omega$ of $T_{AC}(\omega)$  
in Eq.~(\ref{s30a}) and in  Eq.~(\ref{s34b}) we arrive at the relation
\beq
\matel{B} {\bar{b} \; A \, C \,\mbox{$\frac{1\!+\!\gamma_0}{2}$}\,\Gamma\, b(0)
}{B} = \sum_n \; \matel{B} {\bar{b} \, A \,Q(0)}{n} \cdot 
\matel{n} {\bar{Q} \; C \,\Gamma \, b (0)}{B}
\label{s35}
\eeq
which is the intermediate state representation (\ref{s20}). Note that the
projector $(1\!+\!\gamma_0)/2$ in the left hand side can be omitted
since the $\bar b$ field satisfies  $\bar b \!=\! \bar b(1\!+\!\gamma_0)/2$ in
the static limit.

Considering higher values of $k$ in  Eqs.~(\ref{s30a}) and (\ref{s34b})
which describe the subleading in $1/\omega$ terms in the asymptotics of
$T_{AC}(\omega)$, we readily generalize the saturation relation (\ref{s20}):
\beq
\matel{B} {\bar{b} \; A \,\pi_0^k\,\, C \,
\mbox{$\frac{1\!+\!\gamma_0}{2}$}\Gamma\mbox{$\frac{1\!+\!\gamma_0}{2}$}\, b(0)
}{B} = \sum_n \; (E_B\!-\!E_n)^k \,\matel{B} {\bar{b} \, A \,Q(0)}{n} \cdot 
\matel{n} {\bar{Q} \; C \,\mbox{$\frac{1\!+\!\gamma_0}{2}$}\Gamma\mbox{$\frac{1\!+\!\gamma_0}{2}$} \, b (0)}{B}\,.
\label{s36}
\eeq
Thus, each insertion of operator $(-\pi_0)$ inside a composite operator acts as
a factor of the intermediate state excitation energy. This is expected, for
equation of motion of the static quark field $Q$ allows to equate 
$$
i\partial_0 \:\bar{Q} C b(x) =  \bar{Q} \pi_0 C b(x)
$$
for any color-singlet operator $\bar{Q} C b(x)$. At the same time  
$$
i\partial_0 \,\matel{n} {\bar{Q}  C b(x)}{B}
= -(E_n\!-\!M_B)\,\matel{n} {\bar{Q}  C b(x)}{B}.
$$
This reasoning is presented in more detail in Appendix A dedicated to a 
conventional derivation of the saturation relations. 

A couple of comments are in order before closing this subsection. Although we
phrased consideration for the case of expectation values in $B$ mesons at
rest, these assumptions are not mandatory. The very same saturation by complete set 
of heavy quark intermediate states of a given spatial momentum
holds for matrix elements where initial and final states may
be different, and may have nonvanishing momenta. They neither have to be
the ground pseudoscalar states, but with arbitrary spin flavor content.

Likewise, it is worth noting that even the static approximation for $b$ quarks is
actually superfluous; the saturation by physical intermediate states relies solely
upon large mass of their $Q$ quarks. The only modification required for
finite $m_b$ is taking care of projector $\frac{1\!+\!\gamma_0}{2}$ introduced
by the $Q$-quark propagator. Using the identity
$$
1= \mbox{$\frac{1\!+\!\gamma_0}{2}$} + \gamma_5
\mbox{$\frac{1\!+\!\gamma_0}{2}$}\gamma_5 
$$
we arrive at the following generalization: 
\bea
\nonumber
&& \matel{B} {\bar{b} \; A \, C \,\Gamma\, b(0)
}{B}   =  \\
 && \sum_n \; \left(\matel{B} {\bar{b} \, A \,Q(0)}{n} \cdot 
\matel{n} {\bar{Q} \; C \,\Gamma \, b (0)}{B} + 
\matel{B} {\bar{b} \, A \,\gamma_5\,Q(0)}{n} \cdot 
\matel{n} {\bar{Q} \; C \,\gamma_5\Gamma \, b (0)}{B}
\right) . \rule{30pt}{0pt}
\label{s35g}
\eea
The similar relation between $\pi_0$ and the excitation energy is only
modified by the mass shift of the finite-mass $B$ meson:
\bea
\nonumber
\matel{B} {\bar{b} \; A \pi_0^k C \,\Gamma\, b(0)
}{B} &\msp{-5}=\msp{-5}& \sum_n \; [(M_B\!-\!m_b)\!-\!(M_n\!-\!m_Q)]^k \times \\
 &&\msp{-28} 
\left(\matel{B} {\bar{b} \, A \,Q(0)}{n} \cdot 
\matel{n} {\bar{Q} \; C \,\Gamma \, b (0)}{B} + 
\matel{B} {\bar{b} \, A \,\gamma_5\,Q(0)}{n} \cdot 
\matel{n} {\bar{Q} \; C \,\gamma_5\Gamma \, b (0)}{B}
\right) . \rule{30pt}{0pt}
\label{s36g}
\eea

\subsect{Lowest state saturation ansatz}
\label{grstfac}

The intermediate state representation (\ref{s20}) still does not assume any
approximation aside from the static limit for the $b$ quark, yet it may be
used to apply a dynamic QCD approximation. The one we employ here
uses as an input the $B$-meson heavy quark expectation values (\ref{ME}) of
dimension $5$ and $6$, which are expressed  through $\mu_\pi^2$,
$\mu_G^2$, $\rho_D^3$ and  $\rho_{LS}^3$.

All operators with four and more derivatives must have an even number of
spatial derivatives due to rotational invariance. Thus the operators with four
derivatives have either four spatial derivatives, or two time and two spatial
derivatives. Likewise, the $D\!=\!8$ operators with five derivatives may have
four spatial and a single derivative, or two spatial and three time
derivatives. 

We start with the $D\!=\!7$ operators with four spatial derivatives, and
apply (\ref{s35}):
\beq
\matel{B}{\bar{b}\,iD_j iD_k iD_l iD_m \Gamma \,b}{B} \!=\! 
\sum_n \matel{B}{\bar{b}iD_j iD_k b}{n}\,\matel{n}{\bar b \, iD_l iD_m
  \Gamma\, b}{B}. 
\label{s50}
\eeq 
The intermediate states $\state{n}$ in the sum are either the ground-state
multiplet $B, B^*$, or excited states with the suitable parity of light
degrees of freedom.  The ground-state factorization approximation assumes that
the sum in (\ref{s50}) is to a large extent saturated by the ground state
spin-symmetry doublet.  Hence we retain only the contribution of the ground
state and discard the contribution of higher excitations. In the case of
dimension seven operators the result is expressed in terms of the expectation
values with two derivatives, i.e.\ $\mu_\pi^2$ and $\mu_G^2$; matrix elements
involving $B^*$ are related to them by spin symmetry. We illustrate below the
compact result of summation over the multiplet of states. The method we use is
most economic and turns out particularly transparent when generalizing the
ground-state approximation. A derivation using the more conventional
Lorentz-covariant trace formalism is given in Appendix~\ref{appenB}.

Abstracting first from the heavy quark spin we focus on the indices associated
with the light degrees of freedom \cite{rev}; the corresponding ground state
is denoted by $\state{\Omega_0}$ and its (spinor) wavefunction by $\Psi_0$.
The ground-state saturation approximation then reads as
\beq
\matel{\Omega_0}{\bar{Q}\,iD_j iD_k iD_l iD_m \,Q}{\Omega_0} \!=\! 
\matel{\Omega_0}{\bar{Q}iD_j iD_k Q}{\Omega_0}\,\matel{\Omega_0}
{\bar Q \, iD_l iD_m Q }{\Omega_0}
\label{s52}
\eeq 
where the summation over the polarizations of intermediate $\state{\Omega_0}$
is assumed; the spin states of the initial and final $\Omega_0$ may be
arbitrary. Using~\footnote{Note a difference in the definition of
  three-dimensional $\sigma_{mn}$ compared to four-dimensional
  $\sigma_{\mu\nu}$, cf.~Eq.~(\ref{eqmot}).} 
\beq
\matel{\Omega_0}{\bar{Q}\,iD_j iD_k \,Q}{\Omega_0} \!=\! 
\frac{\mu_\pi^2}{3} \Psi_0^\dagger \delta_{jk}\Psi_0 - \frac{\mu_G^2}{6}
\Psi_0^\dagger\sigma_{jk}\Psi_0\,, \qquad \mbox{where~}
\sigma_j \sigma _k \!=\! \delta_{jk}\!+\!\sigma_{jk},
\label{s54}
\eeq 
we get (from now on explicit spinors $\Psi_0^{(\dagger)}$ will be omitted) 
\bea
\matel{\Omega_0}{\bar{Q}\,iD_j iD_k iD_l iD_m \,Q}{\Omega_0} &\msp{-4}=\msp{-4}& 
\frac{(\mu_\pi^2)^2}{9}\delta_{jk}\delta_{lm} - \frac{\mu_\pi^2\mu_G^2}{18}
(\delta_{jk}\sigma_{lm}\!+\!\sigma_{jk}\delta_{lm})
+ \label{s56} \\
&& 
\frac{(\mu_G^2)^2}{36}(
\delta_{jm}\delta_{kl}\!-\!\delta_{jl}\delta_{km}\!+\! 
\delta_{jm}\sigma_{kl}\!-\!\delta_{jl}\sigma_{km}\!+\!
\sigma_{jm}\delta_{kl}\!-\!\sigma_{jl}\delta_{km}), 
\nonumber
\eea 
where spin matrices $\sigma$ act on the spinor indices of the hadron
wavefunctions (heavy quark field $Q$ can be considered spinless). On the
contrary, $\Gamma$ in Eq.~(\ref{s50}) acts on the spin of real $b$ quarks
of QCD; for this reason the corresponding $\sigma$-matrices are denoted by 
$\sigma^Q$ where the confusion is possible. 

The matrix elements for $B$ and $B^*$ mesons in actual QCD, with arbitrary
heavy quark spin matrices $\Gamma^Q$, are directly
expressed through those for the $\Omega_0$-states with spinless heavy
quarks. To obtain them one would take 
the trace over the spin indices of both the heavy quark and of
light degrees of freedom, convoluted with
the corresponding meson spin wavefunctions
\beq
{\cal M}=  B +  \vec B^* \vec \sigma\,,
\label{s58}
\eeq 
where $B$ and $B^*_k$ are the $B$- and $B^*$-meson fields,
respectively. Namely, the generic
matrix element of an operator $\bar{b}\,O\Gamma^Q\,b$ takes the form
\beq
\matel{M}{\bar{b}\, O \Gamma^Q \,b}{M'}=  \frac{1}{2}\, \mbox{Tr}\left[ 
{\cal M}^\dagger \Gamma^Q M' \tilde \Sigma
\right], \qquad \tilde \Sigma\!\equiv \! \sigma_2 \Sigma^T \sigma_2 
\label{s60}
\eeq 
if the matrix $\Sigma$ describes the spinor part of the matrix element of
$Q^\dagger O Q$ between the $\Omega_0$ states, cf.\
Eqs.~(\ref{s52})-(\ref{s56}). The charge-conjugate spinor matrix $\tilde
\Sigma$ is evidently $\Sigma$ itself for $\Sigma\!=\!1$ and $-\Sigma$ for
$\Sigma\!=\!\sigma_k$.

The above rule becomes especially transparent where matrix elements over
spinless $B$ mesons are considered: then $M$, $M'$ are unit matrices and one
simply takes half the trace over spin indices; the structures with an odd
overall number of spin matrices vanish (i.e., spin of the light cloud must
multiply spin of heavy quark in $B$), whereas products of an even number are
reduced to a numeric factor by assuming that $\sigma^Q\!=\!-\sigma$ (total
angular momentum of $B$ meson vanishes!), in agreement with Eqs.~(\ref{s60}).

Using  Eqs.~(\ref{s56}), (\ref{s60}) we get the master equations
for spin-singlet and spin-triplet $B$ expectation values of $D\!=\!7$:
\bea
\frac{1}{2M_B}\matel{B}{\bar{b}\,iD_j iD_k iD_l iD_m \,b}{B} &\msp{-4}=\msp{-4}& 
\frac{(\mu_\pi^2)^2}{9}\delta_{jk}\delta_{lm} + \frac{(\mu_G^2)^2}{36}
\left(\delta_{jm}\delta_{kl}\!-\!\delta_{jl}\delta_{km} \right)
\label{s64}\\
\nonumber
\frac{1}{2M_B}\matel{B}{\bar{b}\,iD_j iD_k iD_l iD_m \,\sigma_{ab}\,b}{B} 
&\msp{-4}=\msp{-4}& 
-\frac{\mu_\pi^2 \mu_G^2}{18}\left(\delta_{jk}\delta_{la}\delta_{mb} \!-\! 
\delta_{jk}\delta_{lb}\delta_{ma}+ \delta_{lm}\delta_{ja}\delta_{kb} \!-\! 
\delta_{lm}\delta_{jb}\delta_{ka}\right) + \\
\nonumber
&&\msp{-1}\frac{(\mu_G^2)^2}{36}\left[
\delta_{jm}(\delta_{lb}\delta_{ka} \!-\!\delta_{la}\delta_{kb}) - 
\delta_{jl}(\delta_{ka}\delta_{mb} \!-\!\delta_{kb}\delta_{ma})+ \right.\\
&&\msp{10}
\left. \delta_{kl}(\delta_{ja}\delta_{mb} \!-\!\delta_{jb}\delta_{ma})-
\delta_{km}(\delta_{ja}\delta_{lb} \!-\!\delta_{jb}\delta_{la})
\right].
\label{s65}
\eea 

The case of the $D\!=\!8$ operators with four spatial and one time derivative
requires minimal modification. Time derivative can occupy the second or fourth
position, or stay in the center, at the third position. In the former case the
evaluation proceeds in the same way, one only needs to complement
Eq.~(\ref{s54}) by the similar relation for Darwin and $LS$ operators:
\beq
\matel{\Omega_0}{\bar{Q}\,iD_j iD_0 iD_k \,Q}{\Omega_0} \!=\! 
-\frac{\rho_D^3}{3} \delta_{jk}-\frac{\rho_{LS}^3}{6} \sigma_{jk} .
\label{s68}
\eeq 
The two related master equations for  $D\!=\!8$ then become 
\bea
\frac{1}{2M_B}
\matel{B}{\bar{b}\,iD_j iD_0 iD_k iD_l iD_m \,b}{B} &\msp{-4}=\msp{-4}& 
-\frac{\rho_D^3\mu_\pi^2}{9}\delta_{jk}\delta_{lm} + \frac{\rho_{LS}^3\mu_G^2}{36}
\left(\delta_{jm}\delta_{kl}\!-\!\delta_{jl}\delta_{km} \right)
\label{s71}\\
\nonumber
\frac{1}{2M_B}
\matel{B}{\bar{b}\,iD_j iD_0 iD_k iD_l iD_m \,\sigma_{ab}\,b}{B} &\msp{-4}=\msp{-4}& 
\frac{\rho_D^3\mu_G^2}{18}\delta_{jk}(\delta_{la}\delta_{mb} \!-\!
\delta_{lb}\delta_{ma})- \\
\nonumber
&&\msp{-50}\frac{\rho_{LS}^3\mu_\pi^2}{18}
\delta_{lm}(\delta_{ja}\delta_{kb} \!-\! \delta_{jb}\delta_{ka}) -
\frac{\mu_G^2\rho_{LS}^3}{36}\left[
\delta_{jm}(\delta_{lb}\delta_{ka} \!-\!\delta_{la}\delta_{kb}) - 
\delta_{jl}(\delta_{ka}\delta_{mb} \!-\!\delta_{kb}\delta_{ma})+ \right.\\
&&\msp{-10}
\left. \delta_{kl}(\delta_{ja}\delta_{mb} \!-\!\delta_{jb}\delta_{ma})-
\delta_{km}(\delta_{ja}\delta_{lb} \!-\!\delta_{jb}\delta_{la})
\right].
\label{s72}
\eea 

For the expectation values with time derivative in the middle position the
corresponding ground-state contribution vanishes; they appear only
due to the `radially' excited states. Therefore, in the ground-state
factorization we set these to zero:
$$
\matel{B}{\bar{b}\,iD_j iD_k iD_0 iD_l iD_m 
\left [\sigma \right ]\,b}{B} 
\stackrel{\scalebox{.55}{\hspace*{-14pt}GSF}}{\Longrightarrow} 0.
$$

Finally, we need to consider the expectation values of the form
$\matel{B}{\bar{b}\,iD_j iD_0^k iD_l \left [\sigma \right ]\,b}{B}$ for
$k\!=\!2,3$ which evidently belong to the tower of $\mu_{\pi,G}^2$ and
$\rho_{D,LS}^3$.  Likewise, their values could be considered as the input
describing strong dynamics, along with the latter; yet they have not been
constrained experimentally.  The intermediate states saturating such
expectation values have opposite parity to the ground state ($P$-wave states)
regardless of number of time derivatives. The counterpart of the ground-state
saturation approximation here is retaining the contribution of the lowest
$P$-wave resonance in the sum; then each power of time derivative amounts to
the extra power of $-\bar \epsilon$, where $\bar
\epsilon\!=\!M_P\!-\!M_B\!\approx \! 0.4\GeV$.

In fact, there are two families of the $P$-wave excitations of $B$ mesons
corresponding to spin of light degrees of freedom $\frac{3}{2}$ or
$\frac{1}{2}$. The combinations $\mu_\pi^2\!-\!\mu_G^2$,
$\rho_D^3\!+\!\rho_{LS}^3$, ... receive contributions only from the 
$\frac{1}{2}$-family, whereas the $\frac{3}{2}$-family gives rise to
$\frac{\mu_\pi^2\!+\!2\mu_G^2}{3}$, $\frac{\rho_D^3\!-\!2\rho_{LS}^3}{3}$,
etc.\ \cite{rev} (the transition amplitude into the 
lowest $\frac{1}{2}$ $P$-state appears
to be suppressed). Therefore, it makes sense to consider these two structures
separately and approximate
\bea 
\matel{B}{\bar{b}\,iD_j (-iD_0)^{k+1}
  iD_l\,b}{B} &\msp{-4}= \msp{-4}&
\left(\bar\epsilon_{3/2}^{\,k}\frac{2\rho_D^3\!-\!\rho_{LS}^3}{9}+
  \bar\epsilon_{1/2}^{\,k}\frac{\rho_D^3\!+\!\rho_{LS}^3}{9}\right) \delta_{jl} \\
\matel{B}{\bar{b}\,iD_j (-iD_0)^{k+1} iD_l \sigma_{jl}\,b}{B} &\msp{-4}=
\msp{-4}& -\bar\epsilon_{3/2}^{\,k}\frac{2\rho_D^3\!-\!\rho_{LS}^3}{3}+
\bar\epsilon_{1/2}^{\,k}\frac{2\rho_D^3\!+\!2\rho_{LS}^3}{3}\,.
\label{s90}
\eea
Note that assuming
$\bar\epsilon_{1/2}=\bar\epsilon_{3/2}=\bar\epsilon$ implies 
$\rho_D^3\!\simeq\!\bar\epsilon \mu_\pi^2$ and 
$-\rho_{LS}^3\!\simeq\!\bar\epsilon \mu_G^2$; the first relation seems to be
satisfied by the preliminary values of $\mu_\pi^2$ and $\rho_D^3$ extracted
from experiment.

\subsect{Summary for ${\cal O}(\Lam^4)$ and  ${\cal O}(\Lam^5)$ expectation
  values}

Combining the above relations we can evaluate all the required nonperturbative
parameters at order $1/m_b^4$ and $1/m_b^5$ in terms of a few quantities
$\mu_\pi^2$, $\mu_G^2$, $\rho_D^3$, $\rho_{LS}^3$ and $\bar{\epsilon}$ taken
as an input.  Tables~\ref{Tab1} and \ref{Tab2} list the resulting expressions
for the expectation values (at
$\bar\epsilon_{1/2}\!=\!\bar\epsilon_{3/2}\!=\!\bar\epsilon$), and give the
corresponding numerical estimates.  For the latter we assume
$\bar\epsilon\!\simeq\!0.4\GeV $, $\mu_\pi^2\!=\!0.45\GeV^2$,
$\mu_G^2\!=\!0.35\GeV^2$, $\rho_D^3\!=\!0.18\GeV^3$,
$\rho_{LS}^3\!=\!-0.12\GeV^3$. \vspace*{5pt}
\begin{table} 
{\small 
\noindent
\begin{tabular}{||c|c|c||c|c|c||c|c|c||} \hline\hline
& expression & $\!\!m_k, \!\scalebox{.85}{$\GeV^4$}\!\!$ & & expression & $\!\!m_k, \!\scalebox{.85}{$\GeV^4$}\!\!$  & &
expression &  $\!\!m_k, \!\scalebox{.85}{$\GeV^4$}\!\!$ \rule[-1pt]{0pt}{10pt} \\ \hline\hline
$m_1$ & $\frac59 \left(\mu_\pi^2\right)^2 $ & $0.11$ &
$m_2$ & $ -\bar \epsilon \rho_D^3 $ & $-0.072$ &
$m_3$ & $  -\frac23 \left(\mu_G^2 \right)^2 $ & $ -0.082$\rule[-5pt]{0pt}{17pt}
 \\ \hline
$m_4$ & $ \!\!\frac43 \left(\mu_\pi^2\right)^2 \!\!+\! \left(\mu_G^2 \right)^2 \!\! $ & $  0.39 $ &
$m_5$ & $  - \bar \epsilon \rho_{LS}^3  $ & $ 0.048 $ &
$m_6$ & $  \frac23 \left(\mu_G^2 \right)^2 $ & $  0.082 $\rule[-5pt]{0pt}{17pt}
 \\ \hline
$m_7$ & $ -\frac83 \mu_\pi^2 \mu_G^2  $ & $ -0.42 $ &
$m_8$ & $ - 8 \mu_\pi^2 \mu_G^2 $ & $ -1.26 $ &
$m_9$ & $  \!\!\left(\mu_G^2 \right)^2 \!\!-\! \frac{10}{3} \mu_\pi^2
\mu_G^2\!\! $ &  $ -0.40  $ \rule[-5pt]{0pt}{17pt}
 \\ 
\hline\hline
\end{tabular}
}
\caption{Expressions and values for the dimension seven matrix elements} 
\label{Tab1} 
\end{table} 
%
%
\begin{table}
{\small 
\noindent
\begin{tabular}{||l|c|c||l|c|c||} \hline\hline
& expression & $\!\!\!r_k, \!\scalebox{.85}{$\GeV^5$}\!\!\!$   & &
expression &  $\!\!\!r_k, \!\scalebox{.85}{$\GeV^5$}\!\!\!$  \\ \hline\hline
$\!r_1\!$ & $ \bar\epsilon^2 \rho_D^3   $ & $   0.029  $ &
$\!r_2\!$ & $  - \mu_\pi^2 \rho_D^3   $ & $  -0.081   $\rule[-5pt]{0pt}{15pt}
 \\ \hline
$\!r_3\!$ & $  -\frac{1}{3}\mu_\pi^2 \rho_D^3  -\frac{1}{6}\mu_G^2
        \rho_{LS}^3   $ & $   -0.020  $ &
$\!r_4\!$ & $ -\frac{1}{3}\mu_\pi^2 \rho_D^3  +\frac{1}{6}\mu_G^2
        \rho_{LS}^3 + \bar\epsilon^2 \rho_D^3  $ & $ -0.005   $\rule[-5pt]{0pt}{15pt}
 \\ \hline
$\!r_5\!$ & $ 0  $ & $ 0   $ &
$\!r_6\!$ & $ \bar\epsilon^2 \rho_D^3  $ & $  0.029   $
 \\ \hline
$\!r_7\!$ & $ 0 $ & $  0  $ &
$\!r_8\!$ & $ \bar\epsilon^2 \rho_{LS}^3  $ & $  -0.019   $\rule[-5pt]{0pt}{15pt}
 \\ \hline
$\!r_9\!$ & $ -\mu_\pi^2 \rho_{LS}^3  $ & $ 0.054   $ &
$\!r_{10}\!$ & $ \mu_G^2 \rho_D^3   $ & $  0.063   $ \rule[-5pt]{0pt}{15pt}
 \\ \hline
$\!r_{11}\!$ & $ \frac{1}{3} \left(\mu_G^2\rho_D^3+\mu_\pi^2 \rho_{LS}^3\right)
                   -\frac{1}{6} \mu_G^2 \rho_{LS}^3  $ & $  0.010   $ &
$\!r_{12}\!$ & $  -\frac{1}{3} \left(\mu_G^2\rho_D^3+\mu_\pi^2 \rho_{LS}^3\right)
                   - \frac{1}{6} \mu_G^2 \rho_{LS}^3  $ & $ 0.004   $ \rule[-5pt]{0pt}{15pt}
 \\ \hline
$\!r_{13\!}$ & $ \!\!\!\frac{1}{3} \left(-\mu_G^2\rho_D^3 \!+\!\mu_\pi^2 \rho_{LS}^3\right)
                   \!+\!\frac{1}{6} \mu_G^2 \rho_{LS}^3 \!\! \! $ & $ -0.046   $ &
$\!r_{14}\!$ & $ \!\! \!\frac{1}{3} \left(\mu_G^2\rho_D^3\!-\!\mu_\pi^2 \rho_{LS}^3\right)
            \!+\!\frac{1}{6} \mu_G^2 \rho_{LS}^3\! +\!\bar\epsilon^2
            \rho_{LS}^3\!\!\! $ & $ 0.013 $ \rule[-5pt]{0pt}{15pt}
 \\ \hline
$\!r_{15}\!$ & $ 0  $ & $ 0   $ &
$\!r_{16\!}$ & $ 0  $ & $  0  $ \rule[-5pt]{0pt}{15pt}
 \\ \hline
$\!r_{17}\!$ & $ \bar\epsilon^2 \rho_{LS}^3  $ & $  -0.019   $ &
$\!r_{18}\!$ & $ 0  $ & $  0  $ \rule[-5pt]{0pt}{15pt}
 \\ 
\hline\hline
\end{tabular}
}
\caption{Expressions and values for the dimension eight matrix elements}
\label{Tab2}
\end{table}


The values in the tables are not precision predictions, of course, for a number of
reasons.  First, they depend on $\mu_\pi^2$,  $\mu_G^2$, $\rho_D^3$,
$\rho_{LS}^3$ which are themselves only known with limited accuracy; the same 
holds true for the value of $\bar\epsilon$. This aspect, however, 
is easy to quantify using the expression in Tables~\ref{Tab1} and \ref{Tab2}. 

Secondly, the estimates are formulated in the infinite mass limit for the 
$b$ quarks, while the parameters of the heavy quark expansion should actually 
include the full mass dependence. 
Generally the finite-mass corrections are governed by the
parameter $\mhad/2m_b$ and can be sizable in $b$ hadrons \cite{chrom}, up to
$15\%$. However, there is a specific suppression of such preasymptotic
correction in the ground-state pseudoscalar mesons related to the observed 
proximity of these states to the so-called `BPS' regime \cite{BPS}. Since we 
deal here exclusively with the pseudoscalar ground state, we expect the finite 
mass corrections to be substantially smaller.

The major issue is thus the validity of the employed approximation for the matrix
elements, in particular retaining only the relevant lowest states.  
The degree to which this ansatz is applicable depends on the operator in question, 
and is expected to deteriorate when the
number of derivatives (i.e. the operator dimension) increases. 

The dominance of the lowest state and suppression of transitions into highly
excited states typically holds for the bound states with a smooth potential. In
field theory for heavy-light mesons this question was studied \cite{lebur} 
in two-dimensional QCD, the so called 't~Hooft model, which is exactly 
solvable in the limit of a large number of colors.  In
particular, the ground-state expectation values for operators with two spatial
derivatives were found to be saturated by the first `$P$-wave' state to an 
amazing degree of accuracy (we should remind that there is no spin in $1+1$
dimensions, therefore only one, not two $P$-wave families). In case this 
also applies to real QCD we would expect a good accuracy of
the employed factorization ansatz for $m_2$ and $m_5$ at order $1/m_b^4$
and a reasonable one for $r_1$, $r_6$, $r_8$, $r_{17}$ (and some other 
related combinations) at order $1/m_b^5$. All such expectation values are
simply given by moments of the combinations of two small-velocity structure
functions $w_{3/2}(\epsilon)$ and $w_{1/2}(\epsilon)$ which are positive,
which strongly constrains the expectation values. 

The situation is a priori less clear with the operators containing four
spatial derivatives. In the 't~Hooft model the effects related to deviations
from the ground-state factorization were studied and were found to be nearly
saturated, again to a very good degree, by the first radial excitation
\cite{lebur}.\footnote{The contribution itself turned out quite significant if 
normalized literally to $(\mu_\pi^2)^2$, apparently since  $\mu_\pi^2$ was
anomalously small there lacking the factor of $3$, the number of space dimensions. 
If normalized to $\La^4$ it was about $3/4$. In actual $B$ mesons $\mu_\pi^2$
is close to $\La^2$.}
For some of the expectation values at order $1/m_b^4$ they can be estimated
following the reasoning of Refs.~\cite{chrom,f0short}. The nonfactorizable
contributions taken at face value appear to be about $50\%$ of the
ground-state one \cite{f0long}.

In reality, the effective interaction in full QCD is singular at short
distances due to perturbative physics. Hard gluon corrections lead to a slow
decrease of the transitions to the highly excited states -- yet they are dual
to perturbation theory. This is taken care of in the Wilsonian renormalization
procedure which is assumed in the kinetic scheme. From this perspective, one
can say that the factorization ansatz yields the expectation values 
at a low normalization point $\mu \!<\! \epsilon_{\rm rad}\!\approx 0.6\GeV$,
before the channels to radially excited states open up; the excitation energy
for such lowest resonance states is probably around $700\MeV$. Of course, the
actual $\mu$-dependence at such low scale does not coincide with the one
derived from perturbation theory; at intermediate excitation energies it must
be in some respect dual to the perturbative one. 

Keeping this in mind it should be appreciated that even for definitely positive
correlators, or those expectation values where all intermediate states
contribute with the same sign, simply adding the first excitation to the
ground state contribution may already constitute some overshooting. Indeed,
a resonance state residing at mass $\epsilon_{\rm rad}$ may be dual to the 
perturbative contribution over the domain of masses 
$$
\frac{\epsilon_{\rm rad}}{2} < \varepsilon < 
\frac{\epsilon_{\rm rad}+ \epsilon_{\rm rad}^{''}}{2}
$$
where  $\epsilon_{\rm rad}^{''}$ is the mass of the second excitation, and
likewise for higher resonances. Then a better approximation for the
expectation value normalized at $\mu\!=\!\epsilon_{\rm rad}$ would be to 
add only a half of the first excitation contribution.
In practical terms, as long as account for the 
power mixing in the perturbative corrections to the conventional Wilson
coefficients (most notably, of the unit operator) has not been extended to
order $1/m_b^4$ and $1/m_b^5$, the effective nonfactorizable piece may turn
out even less.

An additional feature of actual QCD is existence of the 
low-mass continuum contribution beyond
pure resonances, most notably states like $B^{(*)}\pi$ and their $SU(3)$
siblings. Their contribution is typically $1/N_c$ suppressed and usually does
not produce a prominent effect in quantities which are finite in the chiral
limit (and the corrections to factorization for higher-dimension expectation
values are). They can be expected to contribute up to $25\%$ of the ground
state, yet this may be partially offset once the actual QCD conventional
expectation values are used in the factorization ansatz, that likewise
incorporate such states \cite{f0short}.

Considering all these arguments, we  tentatively assign the uncertainty in the
factorization estimate to be at the scale of $50\%$. Yet this should be
understood to apply to the `positive' operators where the lowest-state
contribution does not vanish and the excited state multiplets yield the same-sign 
contribution. The corrections to factorization will be addressed in more
detail in the forthcoming paper \cite{f0long}.

\section{Numerical Estimates for the Higher Order Corrections in the Rate 
and Moments } \label{num} 

Armed with the numerical estimates of all the required expectation values we
are in the position to evaluate the higher-order power corrections to inclusive
$B$ decays. The primary quantity of interest is the total semileptonic width
$\Gamma_{\rm sl}(b\tto c)$ used for the precision extraction of $|V_{cb}|$.

\subsect{$\Gamma(B\tto X_c\,\ell\nu)$}

Assuming the fixed values of $m_b$ and $m_c$ we find the following power
corrections at different orders in $1/m_b$:
\begin{align}
\nonumber
\frac{\delta \Gamma_{1/m^2}}{\Gamma_{\rm tree}} & =  -0.043 & \qquad 
\frac{\delta \Gamma_{1/m^3}}{\Gamma_{\rm tree}} & =  -0.030 & & \\
\frac{\delta \Gamma_{1/m^4}}{\Gamma_{\rm tree}} & =   0.0075 &  \qquad 
\frac{\delta \Gamma^{\rm IC}}{\Gamma_{\rm tree}}& =  0.007 & \qquad 
\frac{\delta \Gamma_{1/m^5}}{\Gamma_{\rm tree}} & =  0.006  \, ,
\label{310}
\end{align}
where $\Gamma_{\rm tree}$ includes the phase space suppression factor of
approximately $0.63$.
We have shown separately the contribution scaling like $1/m_b^3 m_c^2$
and denoted it by $\delta \Gamma^{\rm IC}$. As anticipated \cite{imprec}, it
dominates the high-order effects and may even exceed the $1/m_b^4$
correction, yet it is to some extent offset by the regular $1/m_b^5$ terms. 

The numerical results (\ref{310}) suggest that the power series for
$\Gamma_{\rm sl}(b\tto c)$ is well behaved, and is under good numerical control
provided the nonperturbative expectation values are known. Higher order terms
induce decreasing corrections except where anticipated on theoretical
grounds. The estimated
overall shift due to higher-order terms 
\beq
\frac{\delta \Gamma_{1/m^4}\!+\!\delta \Gamma_{1/m^5}}{\Gamma_{\rm tree}}
\simeq 0.013
\label{312}
\eeq
is well within the interval assessed in \cite{imprec} and, taken at face
value, would yield a $0.65\%\,$ {\sl direct}\, reduction in  $|V_{cb}|$.

This would not be the whole story, however, for the quark masses determining
the partonic width are not known beforehand with an accuracy required to
extract $|V_{cb}|$ with the percent precision. Rather, their relevant
combination is extracted from the fit to the data on kinematic moments of
the  $B\tto X_c\,\ell\nu$ decay distributions, that in turn are affected by
power corrections. 

\subsect{Moments}

The key in the OPE evaluation of $\Gamma_{\rm sl}(b\tto c)$ and, therefore,
in extraction of $|V_{cb}|$ are the first moments of lepton energy
$\aver{E_\ell}$ and of hadron invariant mass squared $\aver{M_X^2}$, which
pinpoint the precision value of the combination of $m_b$ and $m_c$ that
drives the total decay probability. Moreover, analyzed through order $1/m_b^3$
these two moments turned out to depend on nearly the same combination of the
heavy quark parameters. This allowed for a nontrivial cross check of the
OPE-based theory prediction \cite{amst}: essentially, $\aver{M_X^2}$ could be
predicted in terms of $\aver{E_\ell}$ and vice versa, once the heavy quark
parameters were allowed to vary within theoretically acceptable range.

Therefore, besides the practical question about the shift in the fitted heavy quark
parameters, another important issue emerges of whether the consistency
between measured values of $\aver{E_\ell}$ and of $\aver{M_X^2}$ persists once
higher-order corrections are accounted for. 

Numerically we find 
\beq
\delta \aver{E_\ell} = 0.013\GeV, \qquad \delta \aver{M_X^2} = -0.086\GeV^2
\label{314}
\eeq
where the changes shown are a combined contribution of corrections to order
$1/m_b^4$ and $1/m_b^5$. In our analysis we follow Ref.~\cite{slmoments} and
evaluate the moments as the literal ratios 
\beq
\frac{ \int \!{\rm d}E_\ell\, {\rm d}q_0\, {\rm d}q^2  \: K(E_\ell,q_0,q^2)\, C(E_\ell,q_0,q^2) \,
\frac{{\rm d}^3 \Gamma_{\rm sl}}{{\rm d}E_\ell {\rm d}q_0 {\rm d}q^2 }} 
{\int  \!{\rm d}E_\ell \,{\rm d}q_0\, {\rm d}q^2  \: C(E_\ell,q_0,q^2) \,
\frac{{\rm d}^3 \Gamma_{\rm sl}}{{\rm d}E_\ell {\rm d}q_0 {\rm d}q^2 }}
\label{316}
\eeq
without prior expanding the ratio itself in $1/m_b$. Here
$K$ is the corresponding kinematic observable in question (powers of $E_\ell$ or
$M_X^2$) and $C$ is an explicit kinematic cut if imposed. The integrals both in
denominator and in numerator are taken directly as obtained in the OPE through
the corresponding terms in $1/m_b$; in particular, powers of $(M_B\!-\!m_b)$
for hadronic mass moments effectively are {\sl not}\, treated as power
suppressed. Throughout this
paper we perform numeric evaluation of higher-order--induced changes in
observables discarding perturbative corrections altogether; this turned out a
good approximation in the kinetic scheme. 

To assess practical significance of the numerical changes in the moments we 
gauge them considering the amount of commensurate shifts
required 
in the values of
`conventional' heavy quark parameters used in the OPE so far, to make up for 
the new effects. Specifically, we choose to consider $m_b$, $\mu_\pi^2$ and
$\rho_D^3$ for this purpose; changes $|\delta m_b|\!\gsim\!
10\MeV$, $|\delta \mu_\pi^2|\!\gsim\! 0.1\GeV^2$ and  $|\delta
\rho_D^3|\!\gsim\! 0.1\GeV^3$ are deemed significant bearing in mind
the estimated accuracy of the existing OPE predictions \cite{slmoments}.
We do not include $m_c$ here (which implies keeping it fixed throughout the
analysis) for the following reason. The quark mass
dependence of the moments is essentially given by a combination $m_b\!-\!0.7
m_c$; the interval allowed by the fits on individual values of $m_b$ and $m_c$
separately is much wider than $10\MeV$, therefore in practice the required
variation in $m_c$ is not independent and is derived from the corresponding
variation in $m_b$. The stated significance of $\delta m_b \!\sim\!10\MeV$
refers, in fact, to the above combination of masses. 

In actual fits the effect of higher-order corrections is compensated  by a
change in all heavy quark parameters simultaneously. For visualization
purposes we, however, quote for each moment ${\cal M}$ the separate values 
\beq
\delta m_b= -\frac{\delta {\cal M}}{\frac{\partial {\cal M}}{\partial m_b} }
\, ,
\qquad  
\delta \mu_\pi^2=
-\frac{\delta {\cal M}}{\frac{\partial {\cal M}}{\partial \mu_\pi^2} } \, ,
\qquad  \delta \rho_D^3=
-\frac{\delta {\cal M}}{\frac{\partial {\cal M}}{\partial \rho_D^3} }
\label{318}
\eeq
as if only one of them 
were responsible for the adjustment and if the higher-order effect had been made up
for completely. Clearly, if a
particular shift in a heavy quark parameter comes out abnormally large, it
should simply be discarded: this only signals that the moment in question is
insensitive to this parameter and the moment rather constrains other OPE 
parameters. On the contrary, if a shift is small, this generally means that
the parameter is well-constrained and, typically, should not be adjusted. 
In a sense, the values in Eq.~(\ref{318}) would assume that the
quality of the fit before including the calculated corrections has been
perfect, and this clearly is oversimplification. Yet this is suitable to gauge
the significance of the corrections we study. 

The dependence of the considered moments upon heavy quark parameters, entering
denominators in Eq.~(\ref{318}) is given in a ready-to-use form in
Ref.~\cite{slmoments}. From that we obtain
\begin{alignat}{3}
\nonumber
\aver{E_\ell}: \quad &  \delta m_b\!=\! -33\MeV \;(0.022);& \quad  
& \delta \mu_\pi^2 \!=\! -0.39\GeV^2
\;(-0.005);& \quad  & \delta \rho_D^3\!=\! 0.15\GeV^3 \;(0.014) \\
\aver{M_X^2}: \quad & \delta m_b \!=\! -17\MeV \;(0.011); & \quad  
& \delta \mu_\pi^2 \!=\! -0.12\GeV^2
\;(-0.0015);&  \quad  & \delta \rho_D^3\!=\! 0.086\GeV^3 \;(0.008) 
\label{320}
\end{alignat}
The dependence of $\Gamma_{\rm sl}(B)$ on heavy quark parameters has also been
carefully studied \cite{vcb,imprec,slmoments}, and above we have supplemented each
variation by the corresponding relative shift in $|V_{cb}|$, 
\beq
\frac{\delta |V_{cb}|}{|V_{cb}|} = -\frac{1}{2} 
\frac{1}{\Gamma_{\rm sl}}\frac{\partial\Gamma_{\rm sl}}{\partial \,\scalebox{.9}{${\rm HQP}$}}\;
\delta \,\scalebox{.9}{{\rm HQP}} , \qquad
\frac{1}{\Gamma_{\rm sl}} 
\frac{\partial\Gamma_{\rm sl}}{\partial \,\scalebox{.9}{{\rm HQP}}} = \left\{
\begin{array}{ll}
\msp{.3} 0.0013 \MeV^{-1} & \scalebox{.9}{{\rm HQP}}=m_b \\ 
\msp{-4.5} -0.026 \GeV^{-2}& \scalebox{.9}{{\rm HQP}}=\mu_\pi^2\\ 
\msp{-4.5}-0.18 \GeV^{-3} & \scalebox{.9}{{\rm HQP}}=\rho_D^3
\end{array}
\right.
\label{322}
\eeq
the numbers shown in the parenthesis in Eqs.~(\ref{320}). 

Eqs.~(\ref{320}) suggest that one of the possible `solutions' is an increase
in $\rho_D^3$ by about $0.1\GeV^3$; this value may be affected by a variation
in $\mu_\pi^2$ of the scale of $0.05\GeV^2$ or in $m_b$ about $10\MeV$. The
relevance of this solution essentially depends on how precisely the presently
fitted standard heavy quark parameters accommodate both $\aver{E_\ell}$ and 
$\aver{M_X^2}$.

In actuality, the most precise measurements coming from the threshold
production of $B$-mesons at $B$-factories require a lower cut on lepton
energy. Therefore, the proper analysis of the moments should include a
$E_\ell$-cut around $1\GeV$. 

\begin{figure}[h]
\vspace*{7pt}
\begin{center}
\includegraphics[scale=.47]{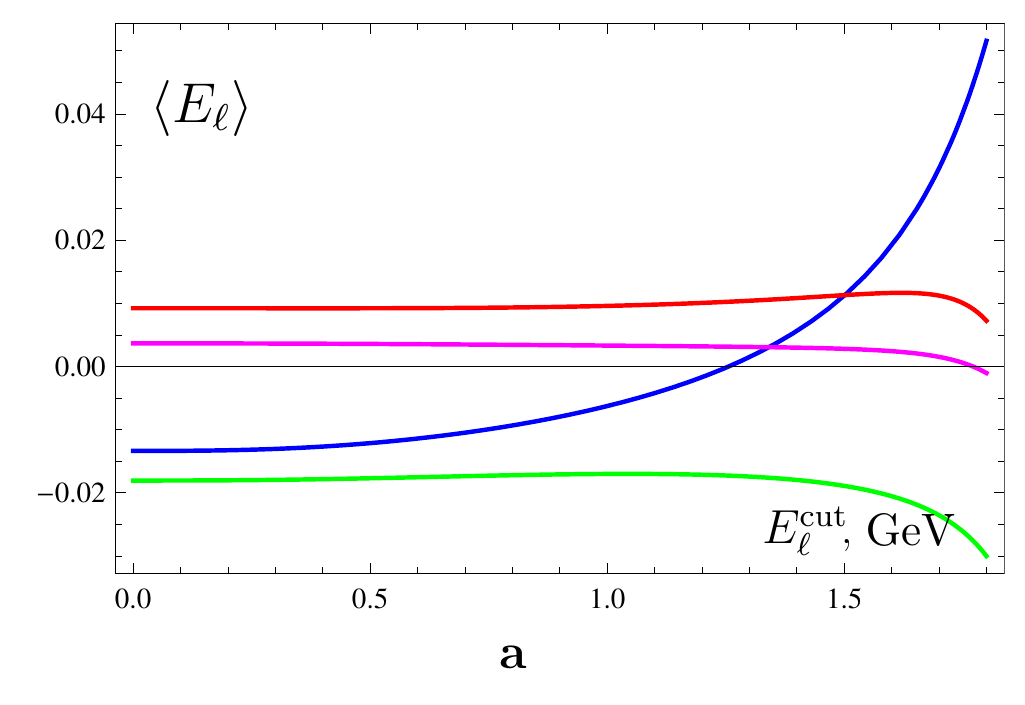}\hfill
\includegraphics[scale=.47]{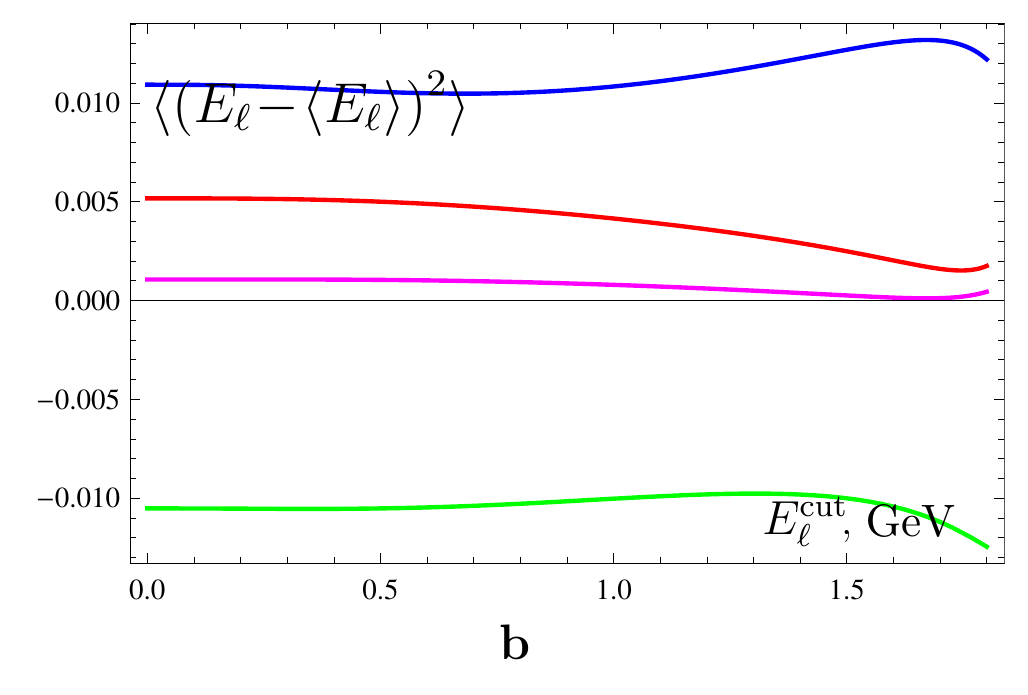}\hfill
\includegraphics[scale=.47]{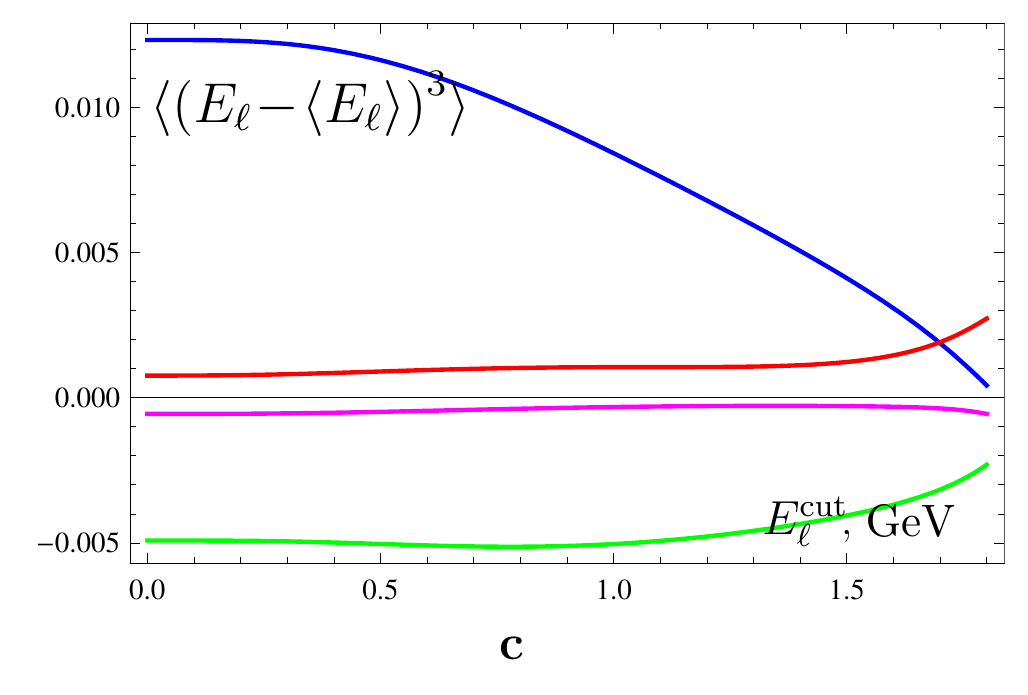}\vspace*{1pt}\\
\includegraphics[scale=.47]{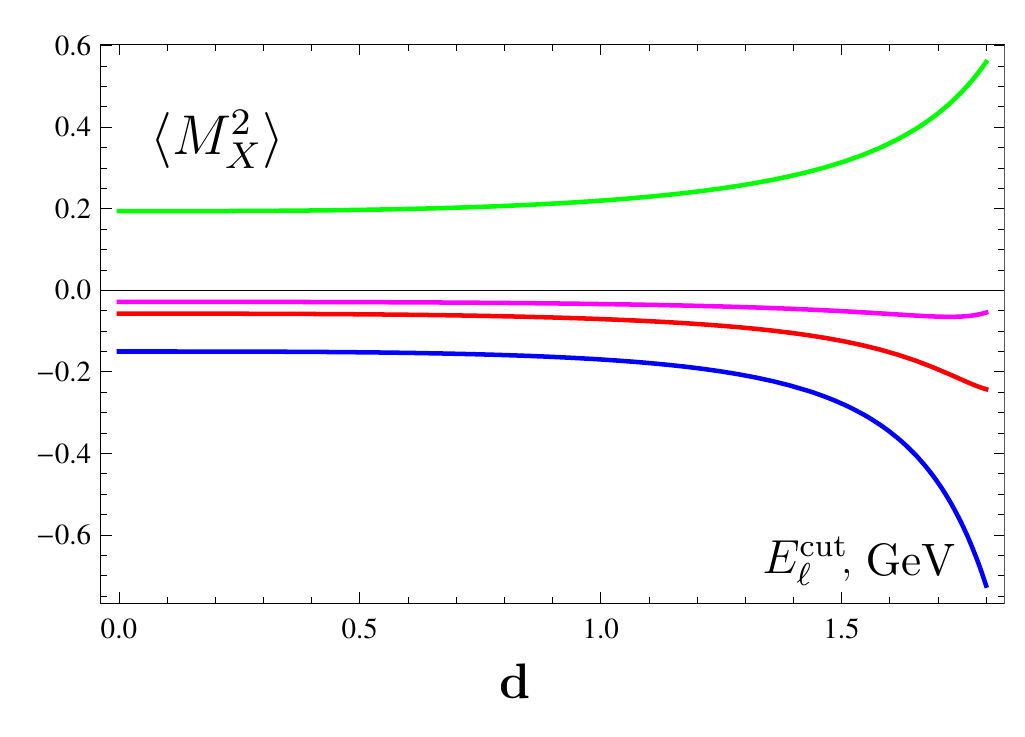}\hfill
\includegraphics[scale=.47]{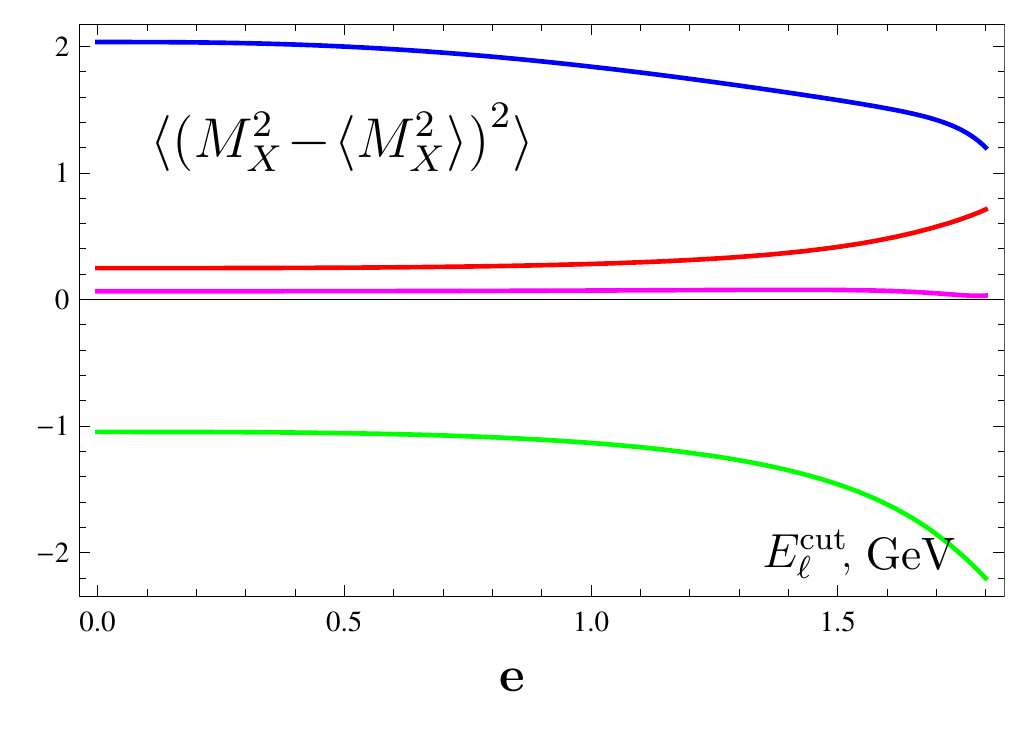}\hfill
\includegraphics[scale=.47]{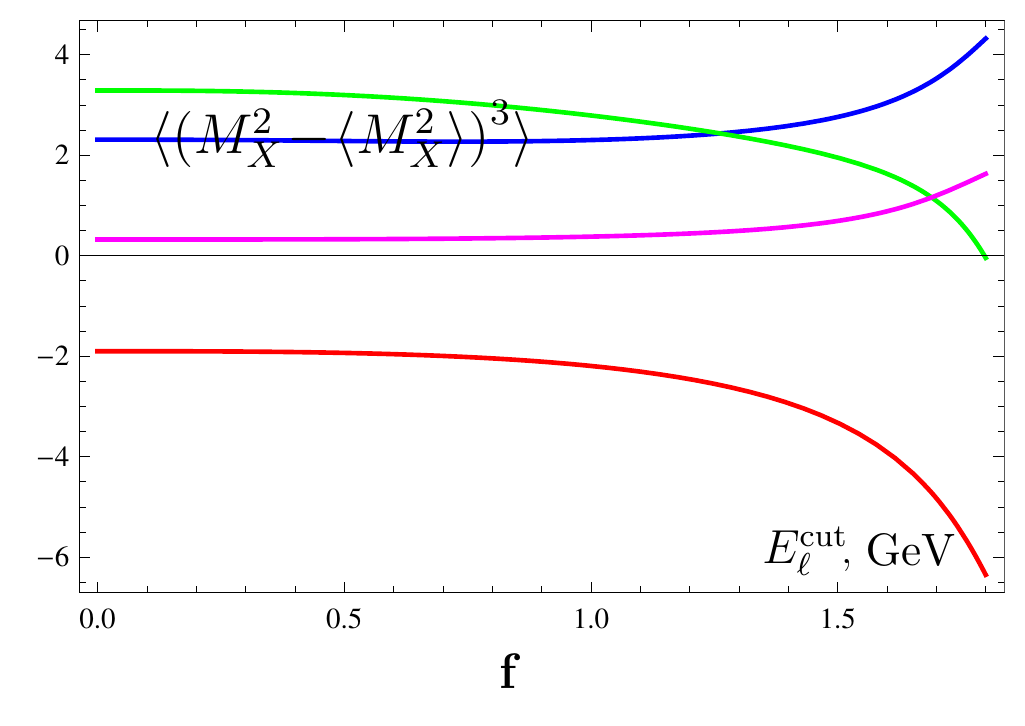}\vspace*{-25pt}
\end{center}
\caption{{\small Power corrections to the first three (central) moments of charged 
lepton energy (upper row) or hadron invariant mass squared (lower row), at
different orders in the $1/m_Q$ expansion, in units of $\GeV$ in the
corresponding power. 
Blue is the effect of $\mu_\pi^2$
and $\mu_G^2$ (order $1/m_b^2$), green at order  $1/m_b^3$, red at
$1/m_b^4$ and magenta finally shows the shift upon including $D\!=\!8$ expectation values
at order $1/m_b^5$}
}
\label{orders}
\end{figure}

Figs.~\ref{orders}\,a--f show the size of the nonperturbative OPE terms for the first three
central lepton energy and hadron mass squared moments depending on the lower
cut $E_\ell^{\rm cut}$ on the charged lepton energy, at different 
orders in $1/m_b$.\footnote{As before, the
explicit factor $M_B\!-\!m_b$ in hadronic mass moments does not count as a
power suppression.}

Keeping in mind that higher moments must generally be sensitive to
higher-order OPE expectation values, we conclude that at moderate cuts
$E_\ell^{\rm cut}\lsim 1.5\GeV$ preserving sufficient `hardness' of the
inclusive probability, the power expansion is well behaved; the $1/m_b^5$
effects are small compared to the $1/m_b^4$ corrections. This is expected
since the IC effects are not parametrically enhanced in the higher moments
\cite{imprec,ic,icsieg}. 

At the same time, it is clear that the estimated effects from higher powers in
the $1/m_b$ expansion are not negligible, in particular in the second and
higher moments. Those are sensitive to 
$\mu_\pi^2$ and $\rho_D^3$, and high-order terms
may produce their sizable shift. 

\begin{figure}[h]
\vspace*{7pt}
\begin{center}
\includegraphics[scale=.5]{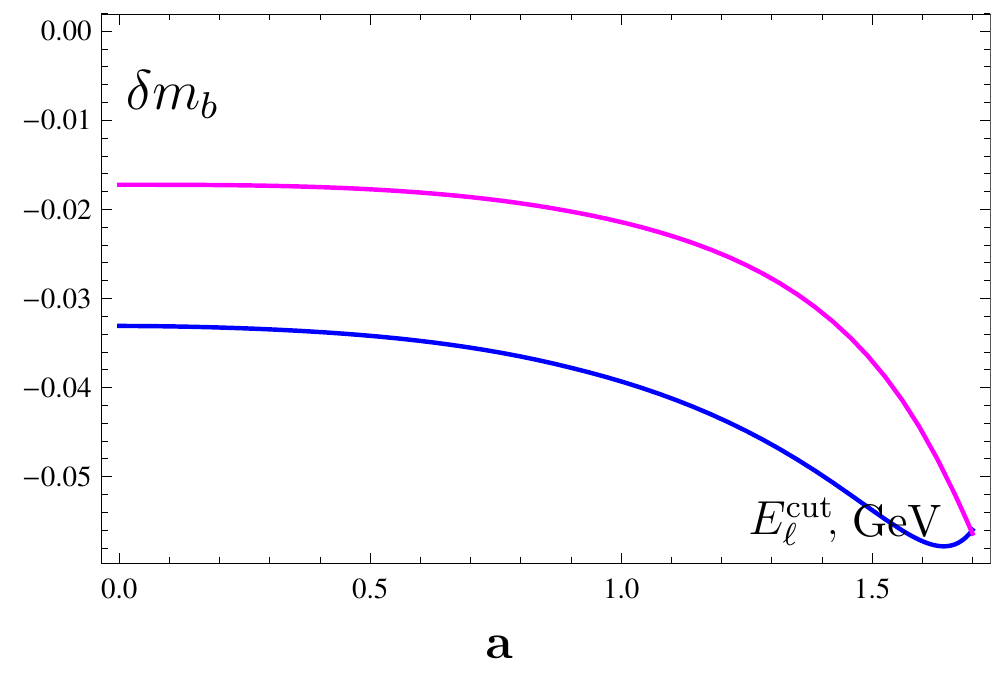}\hfill
\includegraphics[scale=.5]{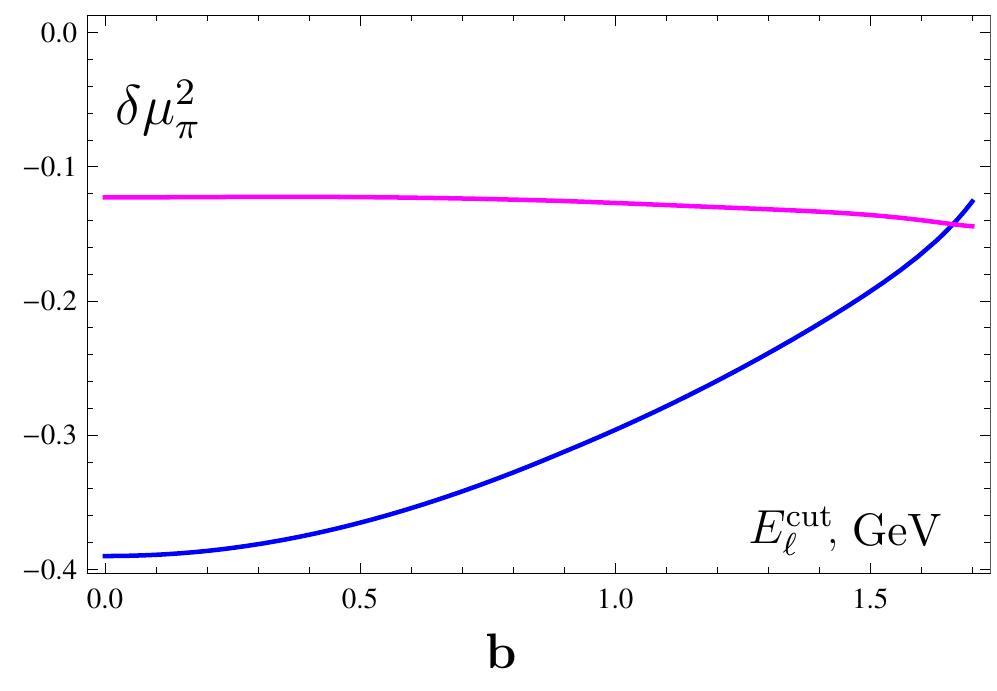}\hfill
\includegraphics[scale=.5]{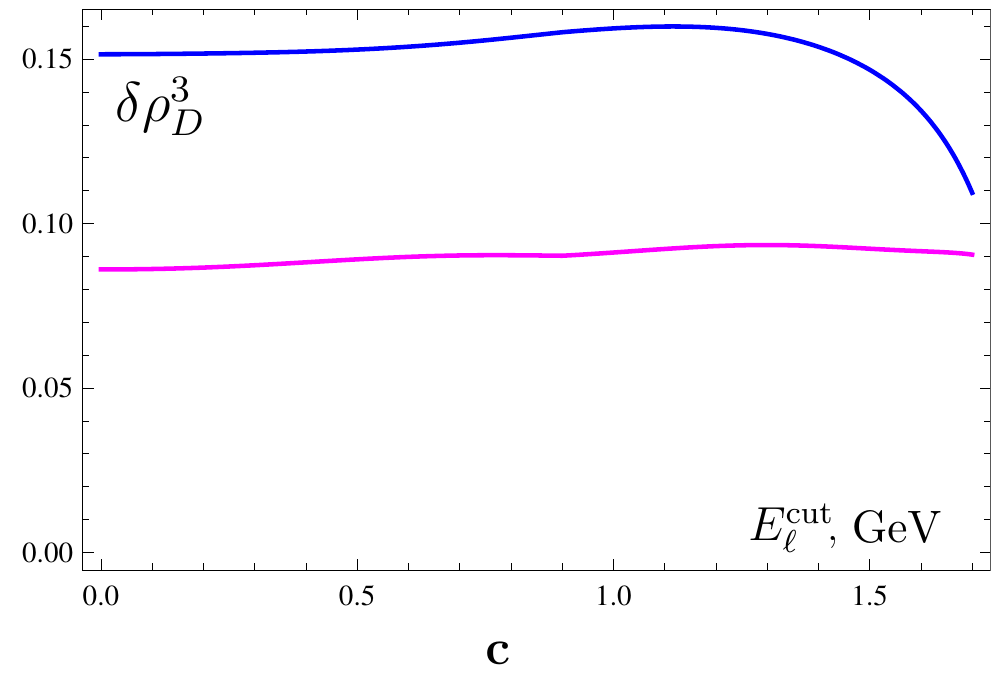}\vspace*{-15pt}
\end{center}
\caption{{\small Changes in $m_b$, in $\mu_\pi^2$ and  in $\rho_D^3$,
    respectively required alone to literally offset the effect of higher-order power 
terms in $\aver{E_\ell}$ (blue) 
and in  $\aver{M_X^2}$ (magenta),  at a given $E_\ell^{\rm cut}$. In units of
$\GeV$ in the corresponding power}
}
\label{l1m2shifts}
\end{figure}

To visualize the potential effect, we have plotted the analogies of the shifts in
Eqs.~(\ref{320}), $\delta m_b(E_\ell^{\rm cut})$, $\delta
\mu_\pi^2(E_\ell^{\rm cut})$, $\delta \rho_D^3(E_\ell^{\rm cut})$ for all six
moments as functions of $E_\ell^{\rm cut}$ and the corresponding $\delta
|V_{cb}|/ |V_{cb}|$. Here we present these dependences only for $\aver{E_\ell}$
and for  $\aver{M_X^2}$, on single plots, separately for $\delta m_b$, 
$\delta \mu_\pi^2$ and $\delta \rho_D^3$, Figs.~\ref{l1m2shifts}\,a--c; the 
special role of these two moments has been mentioned earlier in this subsection.
The corresponding
values (\ref{318}) for all six moments at a representative mild cut 
$E_\ell^{\rm cut}\!= \! 1 \GeV$ are shown in Table~\ref{tableshifts} 
together with the commensurate relative shift $\delta
|V_{cb}|/ |V_{cb}|$. The latter is easily estimated using 
Ref.~\cite{imprec}, cf.\ Eq.~(\ref{322}):
\beq
\frac{\delta |V_{cb}|}{|V_{cb}|} \simeq \left\{
\begin{array}{lll}
\!\! -0.0066 & \mbox{at} & \delta m_b\!=\!10\MeV\\
\,\;0.0013 & \mbox{at} & \delta \mu_\pi^2\!=\!0.1\GeV^2\\
\,\;0.009 & \mbox{at} & \delta \rho_D^3\!=\!0.1\GeV^3
\end{array}
\right.
\label{326}
\eeq

\begin{table}
\begin{tabular}{||c||c|c|c|c|c|c||} \hline\hline
  & \scalebox{.8}{$\aver{E_\ell}$} & \scalebox{.8}{$\aver{(E_\ell \!-\!\aver{E_\ell})^2} $}
& \scalebox{.8}{$\aver{(E_\ell \!-\!\aver{E_\ell})^3} $}
& \scalebox{.8}{$ \aver{M_X^2} $} & \scalebox{.8}{$ \aver{(M_X^2 \!-\!\aver{M_X^2})^2}$}  
& \scalebox{.8}{$ \aver{(M_X^2 \!-\!\aver{M_X^2})^3} $} \\ \hline\hline
$\delta m_b,  \MeV$   & $ -39 $ & $ -60 $ 
&  \raisebox{-7pt}[0pt][0pt]{{\bf ---}}  
& $ -21 $ 
& \raisebox{-7pt}[0pt][0pt]{{\bf ---}} & \raisebox{-7pt}[0pt][0pt]{{\bf ---}} \\
\scalebox{.8}{($\delta |V_{cb}|/|V_{cb}|$)}   & \scalebox{.8}{ ($ 0.026) $} 
& \scalebox{.8}{($0.040$)} & $  $ & \scalebox{.8}{($0.014$)} &   &    \\ \hline
$\delta \mu_\pi^2,  \GeV^2$   & $-0.30$ 
& $-0.12$ & $-0.04$ & $-0.13$
& $-0.08$  & $0.33$\\
\scalebox{.8}{($\delta |V_{cb}|/|V_{cb}|$)}   &  \scalebox{.8}{($-0.004$)} 
&\scalebox{.8}{($ -0.0016$)} & 
\scalebox{.8}{($-0.0005$)} & \scalebox{.8}{($-0.0017$)} & \scalebox{.8}{($-0.0010$)} 
& \scalebox{.8}{($0.0043$)}  \\ \hline
$\delta \rho_D^3,  \GeV^3$   & $ 0.16 $ & $ 0.09 $ & $ 0.02 $ & $ 0.09 $ & $
0.05 $  & $ 0.10 $\\
\scalebox{.8}{($\delta |V_{cb}|/|V_{cb}|$)}   & \scalebox{.8}{($0.014$)} & 
\scalebox{.8}{($0.008$)} & \scalebox{.8}{($0.020$)} & \scalebox{.8}{($0.008$)}
& \scalebox{.8}{($0.005$)} & \scalebox{.8}{($0.009$)}  \\
\hline\hline
\end{tabular}
\caption{Higher order power corrections to the moments with 
$E_\ell^{\rm cut}\!=\!1\GeV$  translated into the required
conventional heavy quark parameter shifts to offset them; also shown are 
relative shifts in
$|V_{cb}|$ these would induce assuming the fixed value of $\Gamma_{\rm
  sl}(B)$. Entries where $\delta m_b$ would exceed $100\MeV$ were left blank}
\label{tableshifts}
\end{table}

As seen from the plots Figs.~\ref{orders}, the cut-dependence of higher-order
corrections shows a generally expected behavior which qualitatively follows
the behavior already found in the $1/m_b^2$ and $1/m_b^3$ effects. The new
corrections likewise show mild cut dependence at $E_\ell^{\rm cut}\!\lsim\! 1
\GeV$ and, typically, sharply change above $E_\ell^{\rm cut}\!\approx \! 1.5
\GeV$, in line with the overall deterioration of the process hardness with
raising cut on $E_\ell$.

\begin{figure}[h]
\vspace*{7pt}
\begin{center}
\includegraphics[scale=.5]{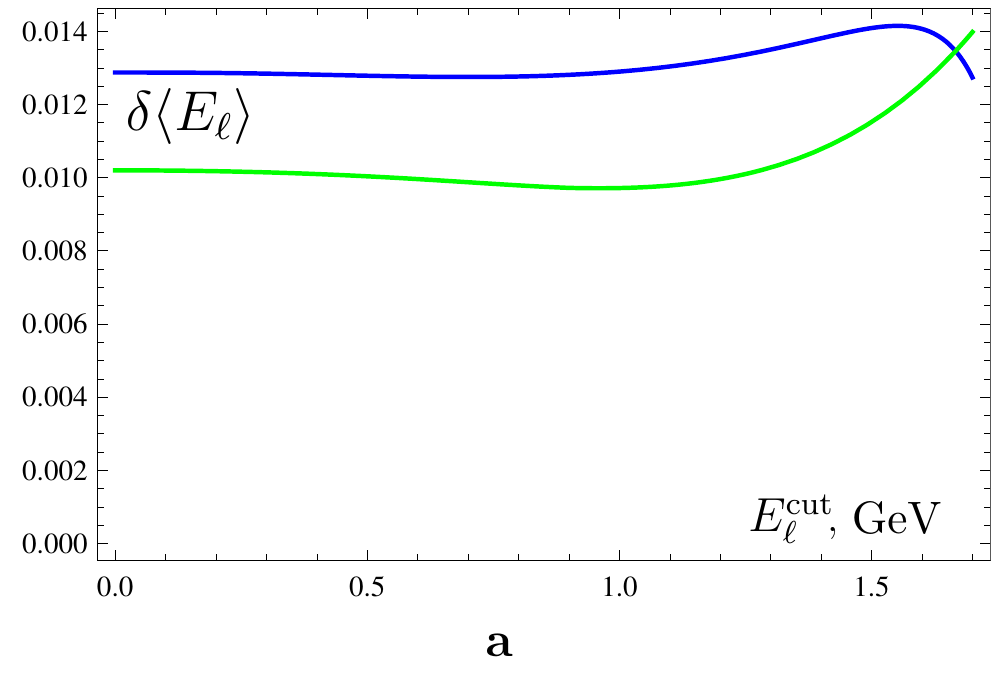}\hfill
\includegraphics[scale=.5]{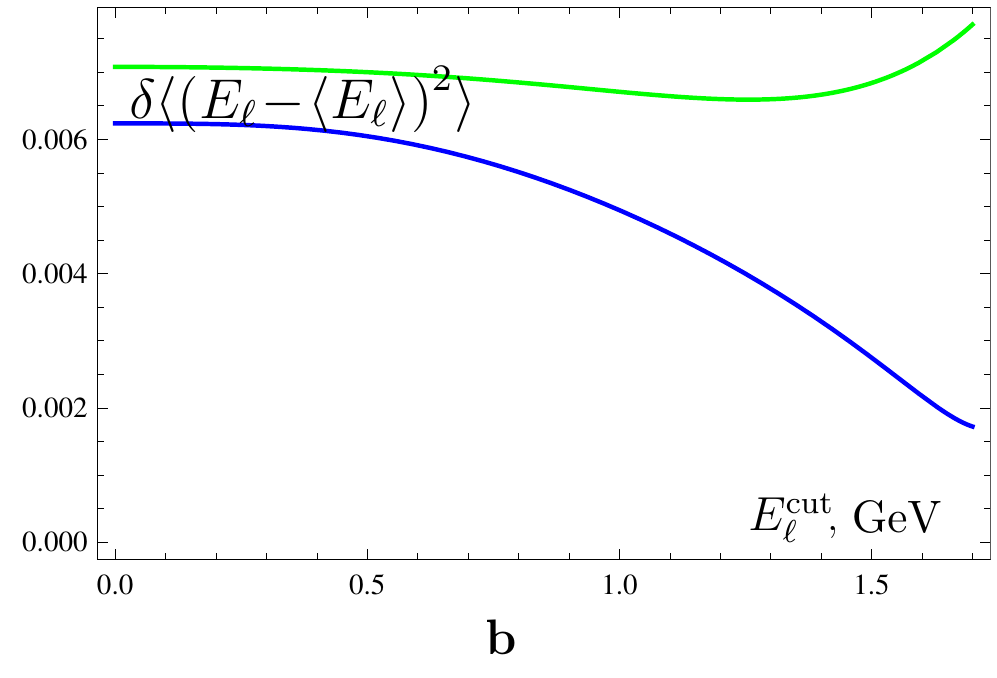}\hfill
\includegraphics[scale=.5]{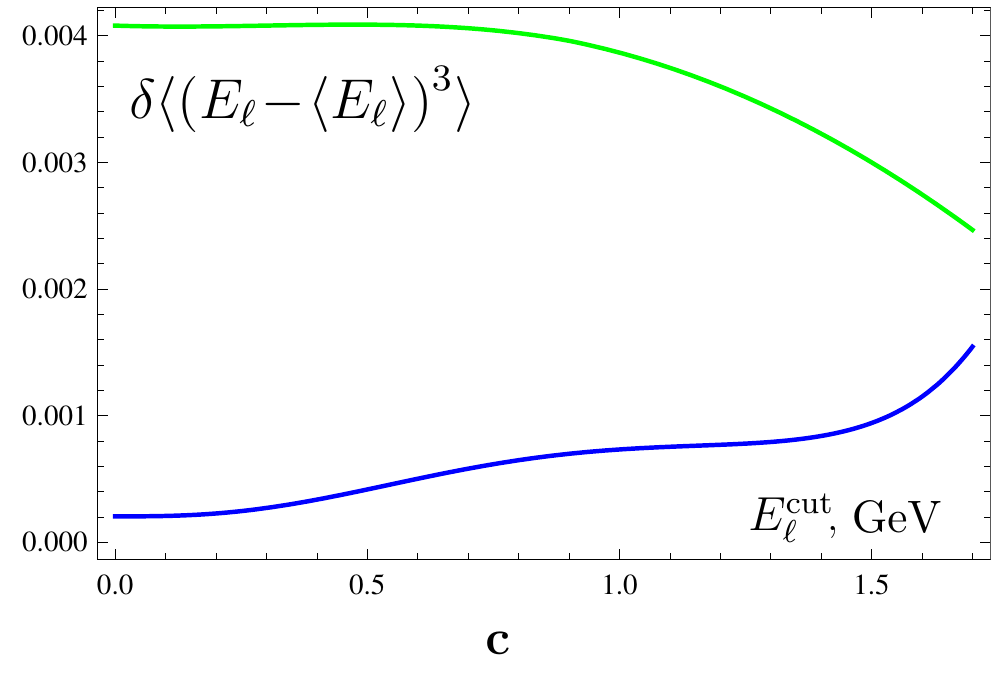}\vspace*{3pt}\\
\includegraphics[scale=.5]{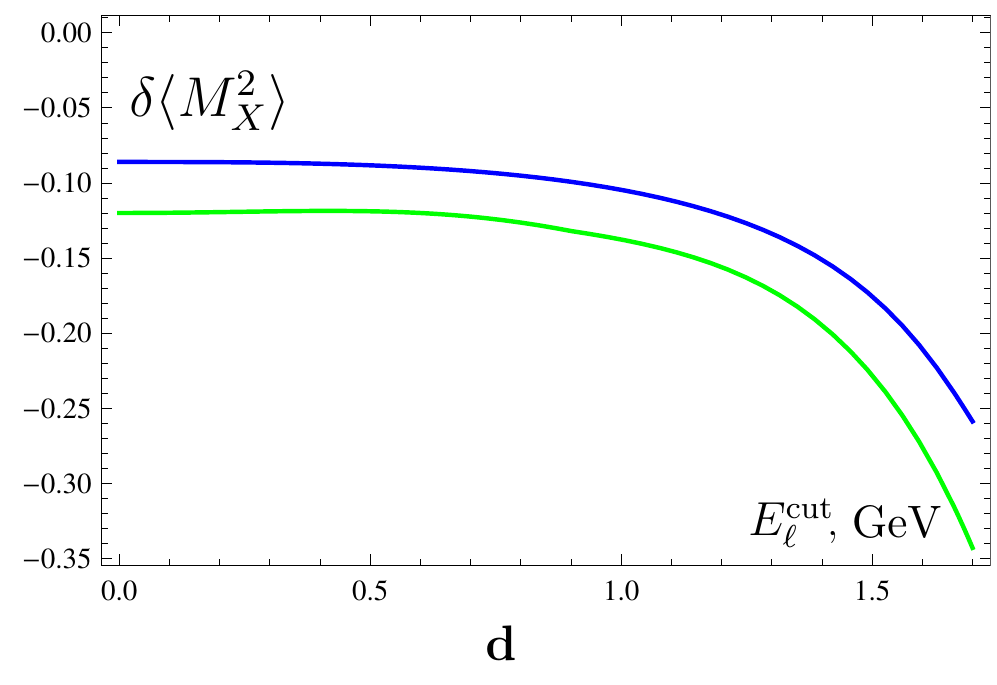}\hfill
\includegraphics[scale=.5]{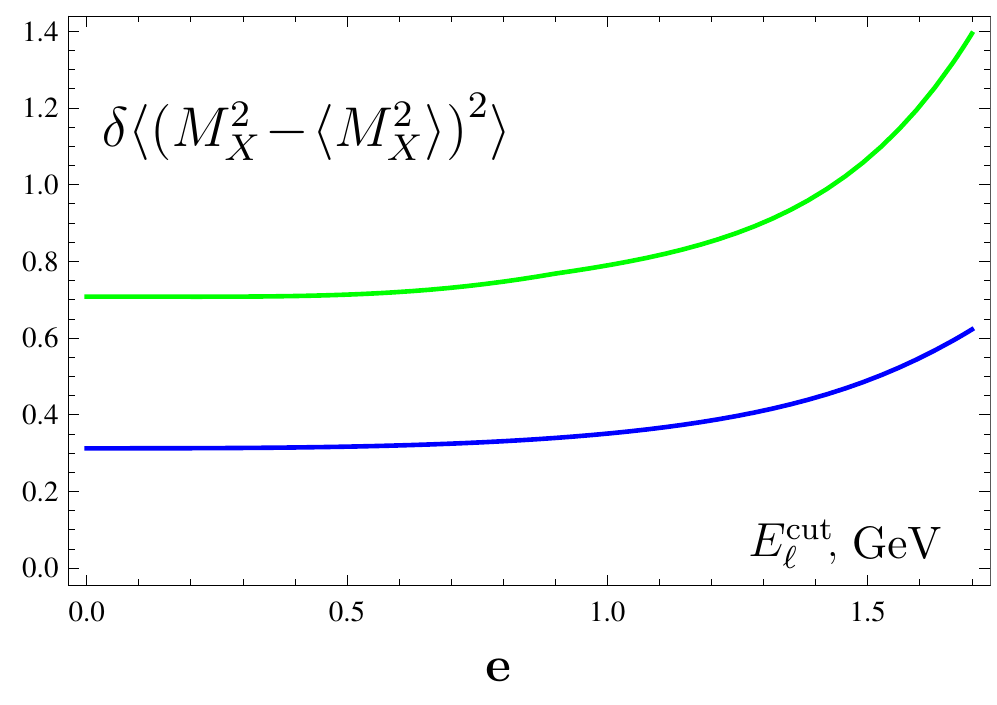}\hfill
\includegraphics[scale=.5]{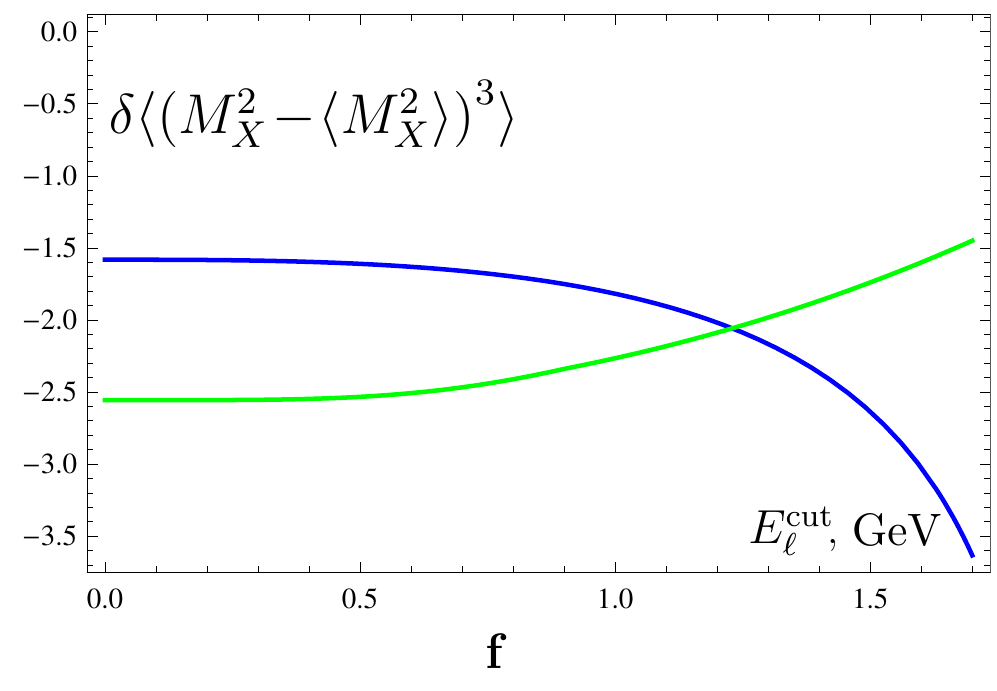}\vspace*{-22pt}
\end{center}
\caption{{\small Effect of including higher-order power 
corrections (blue) to the first three moments of charged 
lepton energy, upper row, and of hadron invariant mass squared, lower row, and 
effect of decreasing the Darwin expectation value by $0.12\GeV^3$ (green), in
$\GeV$ to the corresponding power}
}
\label{shiftdar}
\end{figure}

Based on the numeric pattern of the corrections to the moments we anticipate
that inclusion of higher-order power-suppressed effects will mostly amount to
increase in the fitted value of $\rho_D^3$ by about $0.1\GeV^3$ compared to
the fit where only $D\!=\!5$ and $D\!=\!6$ nonperturbative expectation values
are retained, with a possible shift in $\mu_\pi^2$ by about $\pm 0.05\GeV^2$.
Figs.~\ref{shiftdar}\,a--f illustrate this assertion showing the combined
effect of the new power corrections for the six moments together with the
effect of decreasing $\rho_D^3$ by $0.12\GeV^3$; the similarity of the shifts
suggests that the lack of higher-power corrections in the theoretical
expressions used so far could be to some extent faked by a lower value of the
Darwin expectation value. (The third lepton moment is highly sensitive to the
Darwin expectation value, and uncalculated $\alpha_s$-corrections to the
latter may be blamed for the mismatch apparent in the plot
Figs.~\ref{shiftdar}\,c. Besides, lepton energy moments are to a large extent
saturated by the parton expressions; therefore their high precision allowing
to discuss nonperturbative effects in their value relies on a high degree of
cancellation of conventional perturbative corrections. The extent of such
cancellation at higher loops is not known beforehand, which warrants a
cautious attitude towards theoretical precision of higher lepton moments at
the required level, and places more emphasis in this respect on higher moments
of the hadronic mass.)

A good way to experimentally extract information on the Darwin expectation
value is the third central hadronic mass squared moment. So far $B$-factories
did not attempt to measure it; it has been extracted with an informative
accuracy in the DELPHI analysis \cite{delphi}, however for unspecified reasons
this moment was not included into the global fit, according to HFAG. We find
that the higher-order power corrections (predominantly $1/m_b^4$) tend to
shift this moment by about twice the DELPHI error bar.

Combining the increase in $\rho_D^3$ with the
direct effect on $\Gamma_{\rm sl}$, Eq.~(\ref{312}), we expect an overall
increase in $|V_{cb}|$ by something like $0.4$ percentage points:
$$
\frac{\delta |V_{cb}|}{|V_{cb}|} \approx +(0.003\div 0.005) \,.
$$
This estimate is based on the
expectation values in the ground-state factorization approximation and would
scale with their magnitude; the actual number may be up to a factor of $1.5$
larger. 
We emphasize that this would remain only an educated expectation, for the result
strongly depends on the details of the existing fit to the data, on the precision of
different data points and on their correlations. The final conclusions should be
drawn through incorporating the new corrections in the actual fit to the
data.

\section{Power corrections in  \boldmath $b\tto s+\gamma$  \unboldmath}
\label{bsg}

As another application of the technique described in
Sect.~\ref{gen} we have considered higher-order power corrections
to the decay rate and to the photon energy moments in radiative
$b\tto s+\gamma$ decays. They have been treated in the
approximation of the local weak vertex; no operators with charm
quarks or  chromomagnetic $b\tto s$ vertex were considered. The
analysis parallels that of the semileptonic case, except that the
simplicity of the kinematics (corresponding to $q^2\!=\!0$ in the
latter) leads to reasonably compact analytic expressions even in
higher orders. We have performed the calculations for an
arbitrary mass ratio $m_s/m_b$, however quote here the results
only at $m_s\!=\!0$.

Therefore, we assume the $b\tto s+\gamma$ transition to be
mediated by the effective vertex 
\beq
\frac{\lambda}{2}\, \bar s \sigma_{\mu\nu}(1\!-\!\gamma_5) b \, F^{\mu\nu}.
\label{bsg10}
\eeq
In this approximation power corrections to the integrated decay
rate become
\bea
\nonumber
\Gamma_{bsg}(B)\msp{-5} &=&\msp{-5} \frac{\lambda^2m_b^3}{4\pi} \left[
1-\frac{\mu_\pi^2 \!+ \!3\mu_G^2}{2m_b^2} -
\frac{11\rho_D^3\!-\!9\rho_{LS}^3}{6m_b^3} \right.\\
\nonumber
\msp{-5}& &\msp{-5} \left. +\frac{1}{m_b^4}\left(\frac{1}{8} m_1
\!+ \!\frac{7}{12} m_2\!+ \!\frac{1}{3} m_3\!+\!\frac{7}{8}m_4\!-\! 
\frac{17}{12} m_5\!+\!
\frac{11}{24} m_6\!+\!\frac{1}{32} m_8\!-\!\frac{1}{6}
m_9\right) \right.\\
\nonumber
& &\msp{-4} \left.
\!+
\frac{1}{m_b^5}\left(-\frac{28}{15} r_1 \!-\!\frac{69}{20} r_2\!+\!\frac{21}{20} r_3
\!-\!\frac{107}{60} r_4\!+\!\frac{41}{40}r_5\!+\!\frac{41}{40}r_6\!+\! 
\frac{173}{120} r_7\!+\!\frac{4}{3} r_8\!+\!\frac{31}{24} r_9\right. \right.\\
& &\msp{8} \left.\left.
\!+\!\frac{7}{24} r_{10}\!-\!\frac{13}{8} r_{11}
\!+\!\frac{1}{24} r_{12}\!-\!\frac{1}{6} r_{13}
\!-\!\frac{7}{12} r_{14}\!+\!\frac{17}{24} r_{15}
\!-\!\frac{1}{4} r_{16}\!-\!\frac{3}{8} r_{17}\!+\!\frac{7}{24} r_{18}
\right)
\right]. \qquad
\label{bsg12}
\eea
The first moment, the average photon energy in the decay, is
given by
\bea
\nonumber
2\aver{E_\gamma} \msp{-4} &=&\msp{-4} m_b
+\frac{\mu_\pi^2 \!- \!\mu_G^2}{2m_b} -
\frac{5\rho_D^3\!-\!7\rho_{LS}^3}{6m_b^2} \\
\nonumber
\msp{-4}& &\msp{-4} +\frac{1}{m_b^3}\left(-\frac{1}{8} m_1
\!- \!\frac{23}{12} m_2\!- \!\frac{5}{6} m_3\!+\!\frac{7}{24}m_4\!-\! 
\frac{17}{12} m_5\!+\!
\frac{13}{24} m_6\!+\!\frac{1}{6} m_7\!+\!\frac{3}{32} m_8\!-\!\frac{1}{3}
m_9 \right. \\
\nonumber
& &\msp{8} \left. + \frac{1}{4} (\mu_\pi^2)^2 \!+\!\frac{1}{2} \mu_\pi^2
\mu_G^2 \!-\!\frac{3}{4} (\mu_G^2)^2\right) \\
\nonumber
\msp{-4}& &\msp{-4} 
\!+
\frac{1}{m_b^4}\left(- \frac{5}{4} r_2\!+\!\frac{23}{12} r_3
\!-\!\frac{7}{4} r_4\!-\!\frac{5}{8}r_5\!+\!\frac{39}{8}r_6\!-\! 
\frac{13}{24} r_7\!+\!\frac{25}{24} r_9\!-\!\frac{1}{8} r_{10}\right. \\
\nonumber
& &\msp{8} \left.
\!+ \frac{19}{24} r_{11}\!+\!\frac{7}{24} r_{12}
\!-\!\frac{7}{6} r_{13}\!+\!\frac{19}{12} r_{14}
\!-\!\frac{25}{24} r_{15}\!+\!\frac{7}{12} r_{16}
\!+\!\frac{61}{24} r_{17}\!-\!\frac{17}{8} r_{18}
\right. \\
& &\msp{12} \left.
+\frac{1}{2} \mu_\pi^2 \rho_D^3 \!-\!\frac{13}{6} \mu_G^2 \rho_D^3  
\!+\!\frac{5}{2} \mu_G^2 \rho_{LS}^3\!-\!\frac{1}{6} \mu_\pi^2 \rho_{LS}^3
\right ).
\label{bsg14}
\eea

For the second and third moments we, to simplify the
expressions, quote the corrections to the moments with respect to
$m_b/2$ rather than for the usually considered central moments:
\bea
\nonumber
3\aver{(2E_\gamma\!-\!m_b)^2} \msp{-4} &=&\msp{-4} \mu_\pi^2  -
\frac{2\rho_D^3\!-\!\rho_{LS}^3}{m_b} \\
\nonumber
\msp{-4}& &\msp{-4} +\frac{1}{m_b^2}\left(\frac{3}{4} m_1
\!- \!\frac{7}{2} m_2\!- \! m_3\!+\!\frac{5}{2}m_5 \!+\!
\frac{3}{4} m_6\!-\!\frac{1}{2} m_7\!+\!\frac{11}{16} m_8\!-\!\frac{1}{4}
m_9 \right. \\
\nonumber
& &\msp{8} \left. + \frac{1}{2} (\mu_\pi^2)^2 \!+\!\frac{3}{2} \mu_\pi^2
\mu_G^2 \right) \\
\nonumber
\msp{-4}& &\msp{-4} 
\!+
\frac{1}{m_b^3}\left(-4 r_1 \!+\!5 r_2\!+\!\frac{3}{2} r_3
\!-\!5 r_4\!-\!\frac{21}{4}r_5\!+\!\frac{63}{4}r_6\!-\! 
\frac{23}{4} r_7\!-\!\frac{13}{4} r_9\!+\!\frac{9}{4} r_{10}\right. \\
\nonumber
& &\msp{8} \left.
\!+\!\frac{25}{4} r_{11}\!+\!\frac{21}{4} r_{12}
\!-\!8 r_{13}\!+\! r_{14}
\!+\!\frac{1}{4} r_{15}\!+\!\frac{9}{2} r_{16}
\!+\!\frac{27}{4} r_{17}\!-\!\frac{37}{4} r_{18}
\right. \\
& &\msp{8} \left.
+\frac{5}{6} \mu_\pi^2 \rho_D^3 \!-\! 3 \mu_G^2 \rho_D^3  
\!-\! \mu_\pi^2 \rho_{LS}^3\!+\!\frac{3}{2} \mu_G^2 \rho_{LS}^3
\right ),
\label{bsg16}
\eea
\bea
\nonumber
3\aver{(2E_\gamma\!-\!m_b)^3} \msp{-4} &=&\msp{-4}
-\rho_D^3  +\frac{1}{m_b}\left(\frac{3}{2} m_1
\!- \!2 m_2\!+ \! \frac{1}{4}m_3\!+\!\frac{8}{5} m_5 \!+\!
\frac{1}{10} m_6\!-\!\frac{9}{20} m_7\!+\!\frac{27}{80} m_8\!+\!\frac{1}{10}
m_9  \right) \\
\nonumber
\msp{-4}& &\msp{-4} 
\!+
\frac{1}{m_b^2}\left(- \frac{16}{5} r_1 \!+\!\frac{51}{10} r_2\!+\!\frac{9}{10} r_3
\!-\!\frac{17}{5} r_4\!-\!\frac{27}{10}r_5\!+\!\frac{63}{10}r_6\!-\! 
\frac{8}{5} r_7\!+\!4 r_8\!-\!\frac{11}{5} r_9\right. \\
\nonumber
& &\msp{8} \left.
\!+ \frac{71}{10} r_{10}\!+\!\frac{9}{10} r_{11}
\!+\!\frac{13}{5} r_{12}\!-\!\frac{7}{5} r_{13}
\!-\!\frac{11}{5} r_{14}\!+\!\frac{23}{5} r_{15}
\!+\!\frac{27}{10} r_{16}\!+\!\frac{9}{10} r_{17}\!-\!\frac{37}{10} r_{18}
\right. \\
& &\msp{8} \left.
-\frac{1}{2} (\mu_\pi^2\!+\!3\mu_G^2) \rho_D^3 
\right ).
\label{bsg18}
\eea
The terms through $D\!=\!6$ in Eqs.~(\ref{bsg14}) and (\ref{bsg16})
coincide with the known ones, cf.\ Ref.~\cite{bias}.

Numerical aspects have been analyzed similarly to the semileptonic
moments, employing the factorization approximation of
Sect.~\ref{GSA} for the expectation values. The corrections turn
out rather small not only in the integrated width and in the
average photon energy, but also in the second and even in the
third moments corresponding, in the heavy quark limit, to the
kinetic and Darwin expectation values, respectively. Moreover,
accounting for the $D\!=\!8$ expectation values yields small
effect compared to the  $D\!=\!7$ one, except for the second
moment where both are quite suppressed. Direct evaluation results
in 
\bea
\nonumber
\frac{\delta \Gamma(B\tto X_s\!+\!\gamma )}{\Gamma(B\tto
  X_s\!+\!\gamma)} \msp{-4}& = & \msp{-4}
-0.036_{1/m_b^2}-0.0053_{1/m_b^3}+ 0.00064_{1/m_b^4} +
0.00015_{1/m_b^5} \rule{0pt}{15pt}\\
\nonumber
\delta \aver{2E_\gamma} \msp{-4}& = & \msp{1} 11\MeV_{1/m_b^2}
-14\MeV_{1/m_b^3}+3\MeV_{1/m_b^4} +0.7\MeV_{1/m_b^5}\rule{0pt}{15pt}
\\
\nonumber
12\delta \aver{(E_\gamma\!-\!\aver{E_\gamma})^2}\msp{-4}& = &
\msp{-4} 
-0.106\GeV^2_{1/m_b^3}
+ 0.002\GeV^2_{1/m_b^4}+0.0025\GeV^2_{1/m_b^5} \rule{0pt}{15pt}
\\
24\delta \aver{(E_\gamma\!-\!\aver{E_\gamma})^3}\msp{-4}& = &
\msp{-4} 
\;\;\;\:\, 0.02\GeV^3_{1/m_b^4}-0.0025\GeV^3_{1/m_b^5};\rule{0pt}{15pt}
\label{bsg24}
\eea
the shifts in the second, third and fourth lines here may be
interpreted as an apparent change, up to the sign, in $m_b$,
$\mu_\pi^2$ and $-\rho_D^3$, respectively. 

The small effect on $\mu_\pi^2$ and $\rho_D^3$, far below the
level anticipated in Ref.~\cite{bias}, is somewhat
surprising. The above numbers probably are smaller 
than the corrections due to
the non-valence four-quark operators of the form
$\bar{b}s\,\bar{s}b$ appearing at order ${\cal O}(\alpha_s)$;
they were discussed in Ref.~\cite{bias}, Sect.~4.
It is conceivable that the numerical suppression we obtain 
is partially accidental or is an
artifact of the factorization approximation for the expectation
values. We do not dwell further on this here and plan to look
into the issue in the subsequent studies.

\section{Conclusions and Outlook} 
\label{concl}

In this paper we report a detailed study of higher-order power correction in
inclusive weak decays of heavy flavor hadrons, focused on the semileptonic $B$
meson decays. The calculations existed since the mid 1990s included $1/m_b^2$
and $1/m_b^3$ corrections. We extended the analysis to order $1/m_b^4$ and
$1/m_b^5$; this is still done at tree level, without computing
$\alpha_s$-corrections to the power suppressed Wilson coefficients.

The new calculations contain two elements: deriving the OPE terms proper through the
order in question, and relating the $B$-meson expectation values of the new
operators to a set of hadronic heavy quark parameters of a given
dimension. This has been done in the most general setting, by calculating the
weak decay structure functions of $B$ mesons through order $1/m^5$. They are
expressed in terms of nine new expectation values at order  $1/m^4$ and of
eighteen independent expectation values at order $1/m^5$ (more operators
appear with ${\cal O}(\alpha_s)$ corrections, cf.\ Ref.~\cite{ic}). 
Structure functions
allow one to calculate any inclusive differential distribution incorporating
arbitrary lepton kinematic constraints.

The results for the structure functions are rather lengthy, especially so to
order $1/m^5$, and we do not quote them; they are generated in a computer
program and saved as Mathematica definition files. They are used in this form
to derive the power corrections to various moments of the distributions, both
without and with a cut on the charged lepton energy. 

Since the fast growing number of new heavy quark parameters in successive
orders in $1/m$, such an analysis would be of little practical value without means
to estimate the emerging expectation values with a reasonable confidence. 
We have presented a formal derivation of the saturation representation for the
heavy quark matrix elements of operators encountered in the tree-level OPE
which had been used earlier \cite{ic}, yet often regarded skeptically. Based
on this exact representation a ground-state factorization has been 
formulated and we have presented the resulting expressions explicitly for all
the encountered $B$-meson expectation values. 

These two elements of the higher-order analysis allowed us to arrive at
meaningful numerical estimates and in this way to analyze such important 
questions as the accuracy of the OPE related to the practical truncation of
the $1/m_b$ expansion and the numerical convergence of the OPE series. 

We conclude that the power expansion in inclusive $B$ decays derived from the
OPE is numerically in a good shape as long as the lower cut on lepton energy
remains soft. We find that, generally, the $1/m_b^4$ corrections are somewhat
smaller than those at $1/m_b^3$ and typically tend to partially offset the
latter; the situation is expected to be different for high moments of hadronic
invariant mass. The $1/m_b^5$ terms in usually considered moments are already
noticeably smaller and this suggests that the truncation error at this order
is largely negligible in practice.  The notable exception is the effect of
`Intrinsic Charm' (IC) on the total decay rate which had been argued
\cite{imprec} to be a potentially significant effect being driven, to higher
orders, by an expansion in $1/m_c$ rather than in $1/m_b$.

We have also presented the analytic expressions for the corresponding
higher-order corrections in the $B\tto X_s+\gamma$ decays. A pilot evaluation
suggested that here the effects are numerically insignificant. A
more reliable conclusion, however should include estimates for a wider range of
hadronic parameters.

Based on the pattern of the numerical corrections found for various moments
important in the fit to inclusive semileptonic data, we expect an overall
moderate upward shift in the extracted value of $|V_{cb}|$ about a half
percentage point. This comes as an interplay of the direct downward shift
about $0.65\%$ from the explicit corrections to the semileptonic width and a
larger upward shift due to the change in the heavy quark parameters
extracted from the fit to the moments.

The overall scale of the calculated higher-order power corrections shows that,
as a rule, they are not negligible at the attained level of precision and are
in line with the expectations laid down in Ref.~\cite{slmoments}. Practically
speaking, we expect the $1/m_b^4$ corrections to be of the same scale as the
terms from not yet calculated $\alpha_s$-corrections to the Wilson
coefficients of the chromomagnetic and of the Darwin operators.

We expect that the main effect of the estimated power corrections in the fit
to the semileptonic data will be an increase in the Darwin expectation value
$\rho_D^3$ by about $0.1\GeV^3$, while $\mu_\pi^2$ would not change
significantly; neither the main combination of quark masses $m_b\!-\!0.7m_c$
shaping the mass dependence will change essentially. This does not apply to
$m_b$ or $m_c$ separately; the residual dependence on the absolute values is
subtle and the changes in the central value of $m_b$ as large as $50\MeV$ may
not have real significance. 

An increase in the Darwin expectation value would be welcomed theoretically;
based on the small velocity sum rules and assuming their early saturation we
expect
\beq
\rho_D^3(0.6\GeV)\approx 0.45\GeV \cdot \mu_\pi^2(0.6\GeV)\approx 0.15\GeV^3;
\label{412}
\eeq
evolving this estimate to the routinely used normalization scale $1\GeV$ we
would get a value somewhere around $0.25\GeV^3$.

Extracting the subleading heavy quark parameters from the $E_\ell^{\rm
  cut}$-dependence of the moments, a procedure effectively introduced in the
fits to the data by assuming a strong correlation of theoretical uncertainties
at different $E_\ell^{\rm cut}$ for a given moment, looks an unsafe 
option. The cut-dependence of the moments is rather mild until the cut is
placed relatively high, and there all corrections start to inflate degrading
the theoretical accuracy of the OPE predictions. This is seen in the power
expansion, and the similar behavior is expected from the uncalculated 
perturbative effects. The correction begin to blow up apparently somewhere
near  $E_\ell^{\rm cut}\gsim 1.65\GeV$; to remain on the safe side we suggest
to rely on theory at $E_\ell^{\rm cut}\le 1.5\GeV$.



\vspace*{7pt}

\noindent
{\sl Note added.}\, When the present write-up was being finalized, P.\,Gambino
informed us about a pilot fit to the semileptonic data including
the higher-order corrections to the moments quoted here. The preliminary result
did not follow our expectation that the bulk of the effects reduces to the
increase in $\rho_D^3$, although the suppression of the shift in the extracted
value of $|V_{cb}|$ was observed. We note, however, that the reported outcome
was largely dependent on the assumed strong correlations in the theoretical
uncertainties at different $E_\ell^{\rm cut}$. More reliable conclusions can
be drawn once a more general fit analysis is performed.

\subsection*{Acknowledgments} 
We thank Ikaros Bigi and Paolo Gambino for discussions. This work is 
supported  by the German research foundation 
DFG under contract  MA1187/10-1 and by the German Ministry of Research (BMBF),
contracts 05H09PSF; partial support from the RSGSS-65751.2010.2 grant is
gratefully acknowledged. 


\section{Appendices}

\setcounter{equation}{0}
\renewcommand{\theequation}{A.\arabic{equation}}
\renewcommand{\thetable}{\Alph{table}}
\renewcommand{\thesubsection}{\Alph{subsection}}
\setcounter{section}{0}
\setcounter{table}{0}

\subsect{Conventional derivation of the saturation representation}

Here we present a conventional derivation of the representation of
Eqs.~(\ref{s35}), (\ref{s36}) as a sum over intermediate states, based on the
usual field-theoretic language in the framework of the effective field theory. We
start with the case of operator $A$ without time derivatives:
\beq
\matel{B}{\bar{b}\,A\,C\, b(0)}{B}= \int \!{\rm d}^3\vec x \; \delta^3(\vec{x})\;
\matel{B}{\bar{b}A(0,\vec{x})\,C b(0)}{B};
\label{104}
\eeq
time and space coordinates are explicitly shown separately where
necessary. $A,\,C$ typically contain derivatives; to avoid related ambiguities
one may assume that $A$ acts on the left, on the 
$\bar b$ field.

Using the property of the infinitely heavy quark fields $Q$ in the sector
with a single heavy quark
\beq
Q(0,\vec x) \,\bar{Q}(0)= \delta^3(\vec{x}) \,\mbox{$\frac{1+\gamma_0}{2}$}
\label{106}
\eeq
(which is nothing but to say they are static) we represent the matrix element
as 
\beq
\matel{B}{\bar{b}\,A\,C\, b(0)}{B}= \int \!{\rm d}^3\vec x \;
\matel{B}{\bar{b}\,A Q(0,\vec{x})\: \bar{Q}\,C\, b(0)}{B}. 
\label{108}
\eeq
In this form we can apply the intermediate state 
saturation to the product of two heavy quark operators $\bar{b}\,A
Q(0,\vec{x})$ and $\bar{Q}\,C\, b(0)$:
\beq
\matel{B}{\bar{b}\,A\,C\, b(0)}{B}= \int \!{\rm d}^3\vec x \;
\sum_{n} \matel{B}{\bar{b}\,A Q(0,\vec{x})}{n}\, 
\matel{n}{\bar{Q}C b(0)}{B}.
\label{110}
\eeq
The summation over intermediate states includes both summing over 
internal degrees of freedom for a particular
heavy hadron state (this is traditionally assumed by 
quantum-mechanical summation over states where the 
center-of-mass motion is factored out), and also 
integration over the momentum $\vec{p}$ of the hadronic system as a whole:
\beq
\sum_{n} \equiv \int \frac{{\rm d}^3 \vec p}{(2\pi)^3} \; \sum_{\rm states},
\qquad \state{n} = \state{n_{\scalebox{.5}{\rm QM}}, \vec{p}\,}\,;
\label{112}
\eeq
below we omit the subscript QM when refer to the conventual states with
vanishing total spatial momentum (`quantum-mechanical' states). 

Assuming $B$-meson is at rest, the first transition matrix element in the
r.h.s.\ of Eq.~(\ref{110}) has the following dependence on $\vec{x}$:
\beq
\matel{B}{\bar{b}\,A\, b(0, \vec x)}{n,\vec{p}\,}= e^{i\vec p \vec x}
\matel{B}{\bar{b}A Q(0)}{n}; 
\label{114}
\eeq
the integration over ${\rm d}^3 \vec x\:$
then yields $(2\pi)^3 \delta^3(\vec{p}\,)$ which removes the integration over
the center of mass momentum $\vec p$ of the intermediate state $\state{n}$, and we
arrive at 
\beq
\matel{B}{\bar{b}\,A\,C\, b(0)}{B}= \frac{1}{2M_n}
\sum_{n} \matel{B}{\bar{b}A Q(0)}{n}\, 
\matel{n}{\bar{Q}C b(0)}{B}, 
\label{116}
\eeq
where $\state{n}$ include only the states at rest; the factor $\frac{1}{2M_n}$
appears if one uses their standard relativistic normalization. 

To cover the case of time derivatives we apply the above relation to the
following product:
\beq
\matel{B}{\bar{b}\,A \pi_0^k C\, b(0)}{B}= \frac{1}{2M_n}
\sum_{n} \matel{B}{\bar{b}\,A Q(0)}{n}\, 
\matel{n}{\bar{Q}\,\pi_0^k C\, b(0)}{B}, 
\label{118}
\eeq
where time derivative acts on the right. Now, for any local operator $O$ the
relation holds 
\beq
i\partial_0 \,\matel{n}{O(x_0)}{B}= 
(E_n\!-\!E_0)\matel{n}{O(x_0)}{B},
\label{119}
\eeq
which expresses the fact that the time derivative is given by the commutator
with Hamiltonian. In the case of heavy quark operators one can additionally 
apply equation of motion for the heavy quark field $\bar Q$:
\beq
i\partial_0 \,\matel{n}{\bar{Q} C b(x_0,\vec{0})}{B}= 
\matel{n}{\bar{Q} \stackrel{\rightarrow}{\pi}_0 \!C b(0)}{B}+
\matel{n}{\bar{Q} \stackrel{\leftarrow}{\pi}_0 \!C b(0)}{B}= 
\matel{n}{\bar{Q}\,\stackrel{\rightarrow}{\pi_0} C b(0)}{B}.
\label{120}
\eeq
Combining this with Eq.~(\ref{119}) and repeating, if necessary, this reduction we
obtain 
\beq
\matel{n}{\bar{Q}\pi_0^k C b(0)}{B} = (M_B\!-\!M_n)^k
\matel{n}{\bar{Q} C b(0)}{B}.
\label{122}
\eeq
Therefore, Eq.~(\ref{118}) generalizes relation (\ref{116}) in the following way:
\beq
\matel{B}{\bar{b}\:A\pi_0^k C\, b(0)}{B}= \frac{1}{2M_n}
\sum_{n} \matel{B}{\bar{b}\, A Q(0)}{n}\,(M_B\!-\!M_n)^k\, 
\matel{n}{\bar{Q}\, C\, b(0)}{B},
\label{116g}
\eeq
in agreement with the OPE result  Eq.~(\ref{s36}).



\subsect{Ground-state contribution in conventional Lorentz-covariant trace formalism }
\label{appenB}

Here we derive the expressions for the sum over ground-state heavy quark
symmetry multiplet of meson states, encountered in the factorization
approximation for the $D\!=\!7$ and $8$ expectation values, in the framework
of conventional trace formalism. It was developed in particular to 
describe the heavy quark spin multiplets at different velocities. 
Since only heavy states at rest enter our problem, full
Lorentz covariance is superfluous; of four spinor components used in
conventional formalism only two are independent, corresponding to the spin
indices of $\Omega_0$-states considered in Sect.~\ref{grstfac}.

With the spin of the heavy quark dynamically decoupled in the heavy quark
limit, the $B$ and $B^*$ meson wavefunctions at rest $M$ and
$M^{(*)}_{\lambda}$ can be represented as matrices 
\beq
M \!=\! \sqrt{M_B}\,\frac{1\!+\!\gamma_0}{2} \,i\gamma_5 \qquad 
M^{(*)}_{\lambda}\!=\! \sqrt{M_B}\,\frac{1\!+\!\gamma_0}{2} \,(\vec\gamma
\vec \epsilon_\lambda) ,
\label{m43}
\eeq
\normalmarginpar
where we have equated the $B$ and $B^*$ masses in the heavy quark
limit. One of the indices in these matrices corresponds to the heavy quark
spin and another to that of the light degrees of freedom. Spin symmetry
yields the well known trace formula for the matrix elements:
\beq
\matel{H'}{\bar{b} iD_\mu iD_\nu \Gamma\, b}{H}= 
- {\rm Tr} \left[ \bar M_{H'}\Gamma M_H \Lambda_{\mu\nu}^{\rm light}
\right]
\label{m44}
\eeq
where $H$ and $H'$ are either $B$ or $B^*$, $M_{H, H'}$ are the corresponding
meson wavefunctions and $\Lambda_{\mu \nu}^{\rm light}$ 
encodes dynamics associated with light degrees of freedom. 
While a complicated unknown hadronic tensor in the general
situation, $\Lambda_{\mu \nu}^{\rm  light}$ takes a simple form for 
mesons at rest: 
\beq
\Lambda_{\mu\nu}^{\rm light\,}\rule[-7pt]{.3pt}{17pt}_{\,(v v'\!)=1}=\, 
-\frac{\mu_\pi^2}{3} \Pi_{\mu\nu} + \frac{\mu_G^2}{6}
i\sigma_{\mu\nu}\,;
\label{m45}
\eeq
components other than spatial vanish.
In fact,  $\Lambda_{jk}^{\rm  light}$ amounts to the corresponding 
matrix elements of the heavy quark states $\Omega_0$ of Sect.~\ref{grstfac}.

In order to evaluate the ground-state contribution for the matrix element
(\ref{s50}) we must sum over the states of the $(B,B^*)$ spin-symmetry
doublet. To do this we employ the relation valid for arbitrary $R$
\beq
-\frac{1}{2M_B}  {\rm Tr} \left[
R\bar M\right] M - \sum_\lambda \frac{1}{2M_B} {\rm Tr} \left[
R\bar M^{(*)}_{\lambda}\right] M^{(*)}_{\lambda} =
\frac{1\!+\!\gamma_0}{2} R \frac{1\!-\!\gamma_0}{2}
\label{m46}
\eeq
expressing the completeness of  the $B,\,B^*$ states in the spin space. 
Evaluating the right hand side of (\ref{s50}) in the ground-state
saturation with the trace formula (\ref{m44}) and using relation (\ref{m46}) 
we find 
\bea
\nonumber
\lefteqn {\frac{1}{2M_B}\matel{B}{\bar{b}iD_j iD_k b}{B}\,
\matel{B}{\bar b \, iD_l iD_m
\Gamma \,b}{B}+
\sum_\lambda \frac{1}{2M_B}\matel{B}{\bar{b}iD_j iD_k b}{B^*_\lambda}\,
\matel{B^*_\lambda}{\bar b \, iD_l iD_m  \Gamma\, b}{B}=}\\
\nonumber
& &
\frac{1}{2M_B}  
{\rm Tr} \left[ \bar M M \Lambda_{jk}^{\rm light} \right] 
{\rm Tr} \left[ \bar M \Gamma M \Lambda_{lm}^{\rm light} \right]+
\sum_\lambda 
\frac{1}{2M_B}  
{\rm Tr} \left[ \bar M M^{(*)}_{\lambda} \Lambda_{jk}^{\rm light} \right] 
{\rm Tr} \left[ \bar M^{(*)}_{\lambda} \Gamma M \Lambda_{lm}^{\rm light}
\right]=\\
& &
-{\rm Tr} \left[ \bar M  \frac{1\!+\!\gamma_0}{2} \Gamma M \Lambda_{lm}^{\rm light} 
\frac{1\!-\!\gamma_0}{2}\Lambda_{jk}^{\rm light} \right] = 
-{\rm Tr} \left[ \bar M \Gamma M \Lambda_{lm}^{\rm light} 
\Lambda_{jk}^{\rm light} \right]. 
\label{m48}
\eea
The product of the two hadronic tensors $\Lambda^{\rm light}$ in (\ref{m45}) 
is given by 
\bea
\nonumber
\Lambda_{lm}^{\rm light} \Lambda_{jk}^{\rm light}  
&\msp{-4}=\msp{-4}& 
\frac{(\mu_\pi^2)^2}{9}g_{jk}g_{lm} - \frac{\mu_\pi^2\mu_G^2}{18}
(g_{jk}i\sigma_{lm}\!+\!i\sigma_{jk}g_{lm})
+ \\
&& 
\frac{(\mu_G^2)^2}{36}(
g_{jm}g_{kl}\!-\!g_{jl}g_{km}\!+\! 
g_{jm}i\sigma_{kl}\!-\!g_{jl}i\sigma_{km}\!+\!
i\sigma_{jm}g_{kl}\!-\!i\sigma_{jl}g_{km}).
\label{m49}
\eea 
In the static limit we therefore obtain for spin-singlet and spin-triplet $B$
expectation values 
\bea
\frac{1}{2M_B}  \matel{B}{\bar{b}\,iD_j iD_k iD_l iD_m \,b}{B} &\msp{-4}=\msp{-4}& 
\frac{(\mu_\pi^2)^2}{9}g_{jk}g_{lm} + \frac{(\mu_G^2)^2}{36}
\left(g_{jm}g_{kl}\!-\!g_{jl}g_{km} \right) \qquad
\label{m50}\\
\nonumber
\frac{1}{2M_B}  
\matel{B}{\bar{b}\,iD_j iD_k iD_l iD_m \,(-i\sigma_{ab})\,b}{B} &\msp{-4}=\msp{-4}& 
-\frac{\mu_\pi^2 \mu_G^2}{18}\left(g_{jk}g_{la}g_{mb} \!-\! 
g_{jk}g_{lb}g_{ma}\, + \right. \\
\nonumber
&\msp{-4}& \msp{-62} \left. g_{lm}g_{ja}g_{kb} \!-\! g_{lm}g_{jb}g_{ka}\right) + 
\frac{(\mu_G^2)^2}{36}\left[
g_{jm}(g_{lb}g_{ka} \!-\!g_{la}g_{kb}) - 
g_{jl}(g_{ka}g_{mb} \!-\!g_{kb}g_{ma})\,+
\right.\\
&\msp{-4}& \msp{-9.5}\left.
g_{kl}(g_{ja}g_{mb} \!-\!g_{jb}g_{ma})-
g_{km}(g_{ja}g_{lb} \!-\!g_{jb}g_{la})
\right]\! ,
\label{m51}
\eea 
reproducing Eqs.~(\ref{s64}) and (\ref{s65}).

For the $D\!=\!8$ operators with four spatial and one time derivative we
follow the same route, yet one of the two  $\Lambda^{\rm light}$  now
describes the operator of the form $\bar b iD_j iD_0 iD_k \Gamma
b$. Denoting the  corresponding  hadronic tensor of light degrees of freedom by 
${\cal R}_{\mu\nu}^{\rm light}$,  we express it through the Darwin and Spin-Orbit
expectation values:
\beq
{\cal R}_{\mu\nu}^{\rm light\,}\rule[-7pt]{.3pt}{17pt}_{\,(v v'\!)=1}=
\,\frac{\rho_D^3}{3} \Pi_{\mu\nu} + \frac{\rho_{LS}^3}{6}i\sigma_{\mu\nu} .
\label{m52}
\eeq
Repeating the same steps we end up with
\bea
\frac{1}{2M_B}  
\matel{B}{\bar{b}\,iD_j iD_0 iD_k iD_l iD_m \,b}{B} &\msp{-4}=\msp{-4}& 
-\frac{\rho_D^3\mu_\pi^2}{9}g_{jk}g_{lm} + \frac{\rho_{LS}^3\mu_G^2}{36}
\left(g_{jm}g_{kl}\!-\!g_{jl}g_{km}\! \right) \qquad \qquad
\label{m53}\\
\nonumber
\frac{1}{2M_B}  
\matel{B}{\bar{b}\,iD_j iD_0 iD_k iD_l iD_m (-i\sigma_{ab})\,b}{B} 
&\msp{-4}=\msp{-4}& 
\frac{\rho_D^3\mu_G^2}{18}g_{jk}(g_{la}g_{mb} \!-\!
g_{lb}g_{ma})\, - \\
\nonumber
&\msp{-4}& \msp{-62.3}
\frac{\rho_{LS}^3 \mu_\pi^2}{18}
g_{lm}(g_{ja}g_{kb} \!-\! g_{jb}g_{ka}) 
+ \frac{\mu_G^2 \rho_{LS}^3}{36} 
\left[
g_{jm}(g_{lb}g_{ka} \!-\!g_{la}g_{kb}) - 
g_{jl}(g_{ka}g_{mb} \!-\!g_{kb}g_{ma}) \,+\right.\\
& \msp{-4} & \msp{-2}
\left.
g_{kl}(g_{ja}g_{mb} \!-\!g_{jb}g_{ma})-
g_{km}(g_{ja}g_{lb} \!-\!g_{jb}g_{la})
\right]\!,
\label{m54}
\eea 
identical to  Eqs.~(\ref{s71}) and (\ref{s72}).

\end{document}